\documentclass[usegraphicx,usenatbib]{mn2e}
\usepackage{amssymb}
\usepackage{epsf,multirow}

\usepackage{graphicx}
\def\kms{~km~s$^{-1}$}
\def\micron{$\mu$m}
\def\deg{^{\circ}}
\def\OIII{[O\,{\sc iii}]}
\def\NeV{[Ne\,{\sc v}]}
\def\FeVII{[Fe\,{\sc vii}]}
\def\FeX{[Fe\,{\sc x}]}
\def\FeXI{[Fe\,{\sc xi}]}
\def\FeXIV{[Fe\,{\sc xiv}]}
\def\NeIII{[Ne\,{\sc iii}]}
\def\l{$\lambda$}

\title[Demystifying the CLR of AGNs]{Demystifying the coronal line region of active galactic nuclei: spatially resolved spectroscopy with HST}
\author[Mazzalay et al.]{Ximena Mazzalay$^{1,2}$\thanks{E-mail: ximena@mpe.mpg.de (XM); aardila@lna.br (ARA); skomossa@mpe.mpg.de (SK)}, Alberto Rodr\'iguez-Ardila$^{3}$ and S. Komossa$^{2}$\thanks{Based on observations made with the NASA/ESA Hubble Space Telescope, obtained from the data archive at the Space Telescope Science Institute. STScI is operated by the Association of Universities for Research in Astronomy, Inc. under NASA contract NAS 5-26555.}\\
$^{1}$Instituto de Astronom\'ia Te\'orica y Experimental, CONICET-UNC, Laprida 854, X5000BGR, C\'ordoba, Argentina\\
$^{2}$Max-Planck-Institut f\"ur extraterrestrische Physik, Postfach 1312, 85741 Garching, Germany\\
$^{3}$Laborat\'orio Nacional de Astrof\'isica, Rua dos Estados Unidos 154, Bairro das Nac\"oes, CEP 37500-000, Itajub\'a, MG, Brazil
}

\begin{document}

\date{}


\maketitle


\begin{abstract}
We present an analysis of STIS/HST optical spectra of a sample of ten Seyfert galaxies aimed at studying the structure and physical properties of the coronal-line region (CLR). The high-spatial resolution provided by STIS allowed us to resolve the CLR and obtain key information about the kinematics of the coronal-line gas, measure directly its spatial scale, and study the mechanisms that drive the high-ionisation lines. 
We find CLRs extending from just a few parsecs ($\sim 10$~pc) up to 230~pc in radius, consistent with the bulk of the coronal lines (CLs) originating between the BLR and NLR, and extending into the NLR in the case of \FeVII\ and \NeV\ lines. The CL profiles strongly vary with the distance to the nucleus. We observed line splitting in the core of some of the galaxies. Line peak shifts, both red- and blue-shifts, typically reached 500\kms, and even higher velocities (1000\kms) in some of the galaxies. In general, CLs follow the same pattern of rotation curves as low-ionisation lines like \OIII. 
From a direct comparison between the radio and the CL emission we find that neither the strength nor the kinematics of the CLs scale in any obvious and strong way with the radio jets. Moreover, the similarity of the flux distributions and kinematics of the CLs and low-ionisation lines, the low temperatures derived for the gas, and the success of photoionisation models to reproduce, within a factor of few, the observed line ratios, point towards photoionisation as the main driving mechanism of CLs. 
\end{abstract}


\section{Introduction}\label{intro}

Emission-line spectroscopy of Active Galactic Nuclei
(AGN) provides a wealth of information
on the gaseous components in galaxy cores,
and has been employed since the beginning of
the twentieth century (e.g., Fath 1909).
It allows us to study the
physical conditions, gas kinematics, metal abundances,
and their redshift evolution, provides us with means of estimating
black hole masses in AGN, and has enabled us to
study the evolution of galaxy -- black hole scaling
relations out to high redshift.
Apart from the classical broad-line region (BLR) at small
core distances, and the extended classical narrow-line region (NLR),
a subset of AGN spectra show lines from very highly
ionised atoms, known as ``Coronal lines'' (CLs) because they were
first observed in the solar corona. These lines are collisionally
excited forbidden transitions
within low-lying levels of highly ionised species with
ionisation potentials $\rm{IP} \geq 100$~eV. In AGNs, the CLs are emitted in the so-called coronal
line region (CLR), likely located outside
the bulk of the BLR, and inside the bulk of the NLR.

CLs have been detected in the optical
and infrared spectra of all types of Seyfert galaxies,
including type 1s, type 2s, and narrow-line Seyfert 1 galaxies
(e.g., Seyfert 1943; Penston et al. 1984; Marconi et al. 1994; Nagao, Taniguchi \& Murayama 2000; Sturm et al. 2002; Rodr\'iguez-Ardila et al. 2002, 2006; Deo et al. 2007; Mullaney \& Ward 2008; Komossa et al. 2008; Gelbord, Mullaney \& Ward 2009) and also radio galaxies (e.g., Best, R\"{o}ttgering \& Lehnert  1999; Holt, Tadhunter \& Morganti 2006), and represent one of the key gaseous components of the active nucleus.
They appear to be approximately equally abundant in type 1
and type 2 AGN (Rodriguez-Ardila et al., in prep).

Since the mid 70s, some evidence has been reported that optical
CLs tend to be broader than low-ionisation forbidden lines
(Phillips \& Osterbrock 1975; Cooke et al. 1976) and that
their centroid position is blueshifted with respect to the systemic velocity
of the galaxy (Grandi 1978; Penston et al. 1984). This
has been interpreted as evidence for a location of the CLR
closer to central engine than the classical
NLR and probably associated with outflows (e.g., Ward \& Morris
1984; Mullaney et al. 2009).
Consistent with this scenario is the correlation found
between the ionisation potential necessary to create the
ionised species and the line width, seen in some (but not in all)
Seyfert galaxies (Wilson 1979; Pelat, Alloin \& Fosbury 1981; Evans
1988; Erkens, Appenzeller \& Wagner 1997).

So far, the precise nature and origin of CLs remained
uncertain. Several models have been considered in explaining
the CLs, including winds from the molecular torus
(e.g., Pier \& Voit 1995; Nagao et al. 2000; Mullaney et al. 2009),
an origin within the (X-ray) ionised absorbers
(e.g., Komossa \& Fink 1997a, b;
Porquet et al. 1999), a high-ionisation component of the inner NLR (e.g., Komossa \& Schulz 1997; Ferguson, Korista \& Ferland 1997b; Binette et al. 1997), 
and a low-density component of the ISM (Korista \& Ferland 1989).

In most observational studies, the CLs are not directly spatially
resolved, due to the fact that the CLR is more compact than the classical NLR.
One indirect way to obtain spatial information is by variability studies.
Very few galaxies have ever shown dramatic variability of their CLs
[IC\,3599 (Brandt, Pounds \& Fink 1995; Grupe et al. 1995; Komossa \& Bade 1999)
and SDSSJ0952+2143 (Komossa et al. 2009)], likely a response
to a rare high-amplitude outburst in the
ionising radiation from a temporary accretion event.
Mild CL variability is occasionally
seen (e.g., Netzer 1974; Veilleux 1988),
but as a class, CLs do not vary very much at all \citep{Veilleux}.

A more direct way to resolve the CLR has become possible only recently by means
of Hubble Space Telescope (HST) and ground-based integral-field spectroscopy.
Recent advances in the understanding of the CLR include the
determination of the size and morphology of high-ionisation gas by means
of infrared HST and ground-based AO infrared imaging/spectroscopy in a few objects
(e.g., Thompson et al. 2001; Prieto, Marco \& Gallimore 2005; S\'anchez et al. 2006; Riffel et al. 2008; Bedregal et al. 2009; Storchi-Bergmann et al. 2009). The results indicate CLRs
with sizes varying from compact ($\sim$30~pc) to extended ($\sim$200~pc)
and the CLR is aligned preferentially with the direction of the lower ionisation
cones seen in these sources. Moreover, the most highly ionised species
show the smallest sizes of the emitting regions. In the optical regime, HST spectroscopy was employed to study spatially resolved emission-line ratios of nearby Seyfert galaxies, including in the measurements and models the lowest-ionisation coronal lines \NeV\ and \FeVII\ 
(e.g., Nelson et al. 2000; Kraemer \& Crenshaw 2000a; Kraemer et al. 2000; Whittle et al. 2005; Collins et al. 2005, 2009).

The study of CLs is important for several reasons.
Firstly, CLs with their high-ionisation potentials
trace an important part of the ionising
continuum in the EUV to soft X-ray regime, 
which is not always directly
accessible from observations. Secondly, the CL \NeV\l3426 is often the only forbidden line
identified in high-redshift spectra of AGN
and is therefore used to
identify AGN in deep and wide multi-wavelength surveys.
In the future, the line properties (flux and profile) may be further
used as a diagnostic of the AGN properties, potentially including
a measurement of host galaxy velocity dispersion.
A better understanding of the site and conditions of formation of this
line is therefore of great interest.
In several galaxies, there is strong evidence that the bulk of the CLR 
is in outflow (e.g., Erkens et al. 1997; Mullaney et al. 2009).
If this is a global property of CLRs in all AGN, their study provides
us with new constraints on the formation and driving
mechanism (e.g., radiation pressure versus
lateral flows around nuclear radio jets)
of such outflows in the very cores of AGN. Observations
of high-spatial resolution are thus essential
in spatially resolving the scales close to where these
outflows might form. The widths, profiles, and
wavelength shifts of the CLs constitute important diagnostics of
these energetic processes (such as outflows, or also shocks)
in regions close to the central engine.

Given that most previous CLR studies were based on seeing-limited
ground-based spectroscopy, key to making further progress in understanding the CLR, and to testing CLR models is spatially resolved spectroscopy,
which directly zooms into the central tenths of parsecs of the nearby
Seyfert galaxies. 
The highest possible spatial resolution to date
can only be achieved with observations above the atmosphere,
as they are possible with the HST.
Here, we present Space Telescope Imaging
Spectrograph (STIS) optical spectroscopy of a selected sample of
local AGNs with the goals of studying the structure and physical
properties
of the CLR. The spectra, described in Section 2, allow
us for the first
time the simultaneous study of coronal lines with $100 < \rm {IP} <
504$~eV in a spatially resolved fashion. Questions
that we will address are the extent of the CLR (Section 3),
its kinematics and how it compares to that of the low-ionisation gas
(Section 4), and the role of shocks and photoionisation in determining the strengths of CLs 
(Section 5). The main conclusions of this work are summarised in Section 6.

\section{Sample selection, observations, and data reduction}

We retrieved archival optical HST/STIS spectra of 10 Seyfert galaxies listed in column 1 of 
Table~\ref{properties}. The original sample was composed of 101 Seyfert galaxies observed 
with STIS and published by Spinelli et al. (2006). Our selection was
based on two criteria: first, we chose galaxies 
with previous coronal line detections and second, we restricted the sample to relatively 
nearby objects ($z<0.018$) in order to obtain high-spatial resolution in physical scales. In all cases,
the detector consisted of a 1024 x 1024 CCD with a spatial scale of 0.051 arcsec/pixel, 
giving spatial resolutions ranging from 3 to 17~pc (for an assumed value of 
H$_0=75$~km~s$^{-1}$~Mpc$^{-1}$). 

Since the galaxies of our sample were observed in different STIS observing programs, the 
spectra were obtained in different instrumental configurations. In Table~\ref{Observations} we present a log of these 
observations. The spectral region around the H$\beta$ emission line was covered by the G430L grating (2900--5700~\AA, hereafter blue spectra), with a spectral resolution of 2.73~\AA/pix. When available, the region around the H$\alpha$ emission line was covered either by the G750L (5250--10300~\AA) or the G750M (6300--6850~\AA) grating (hereafter red spectra), with spectral resolutions of 4.92~\AA/pix and 0.56~\AA/pix, respectively.

The retrieved 2D spectra were already rectified and wavelength and flux calibrated. The images
were converted from surface brightness per Angstrom 
(in erg~cm$^{-2}$~sec$^{-1}$~\AA$^{-1}$~arcsec$^{-2}$) to flux per Angstrom (in erg~cm$^{-2}$~sec$^{-1}$~\AA$^{-1}$) 
following the procedure described in section 5.4.1 of the STIS Data Handbook\footnote{http://www.stsci.edu/hst/stis/documents/handbooks/}. 
Some of the galaxies included multiple exposures, dithering the target along the slit 
(see Table~\ref{Observations}). We shifted and combined these images using the imshift and imcombine 
tasks of {\sc iraf}, respectively.

\begin{table}
\caption{Properties of the galaxies in the sample.}
\label{properties}
\begin{tabular}{lccc}
\hline
Object    & Seyfert  & $z$      & Scale \\
	  & Type     &          & [pc~arcsec$^{-1}$]\\
\hline
Mrk~3    & Sy2      & 0.01350 & 260 \\
Mrk~348  & Sy2	    & 0.01503 & 290 \\
Mrk~573  & Sy2      & 0.01717 & 330 \\
NGC~1068 & Sy2      & 0.00379 & 73  \\
NGC~3081 & Sy2      & 0.00797 & 155 \\
NGC~3227 & Sy1.5    & 0.00385 & 75  \\
NGC~4151 & Sy1.5    & 0.00331 & 65  \\
NGC~4507 & Sy1.9    & 0.01180 & 230 \\
NGC~5643 & Sy2      & 0.00399 & 77  \\
NGC~7682 & Sy2      & 0.01714 & 330 \\
\hline

\end{tabular}
\end{table}

\begin{table*}
\begin{minipage}{126mm}
 \caption{Observation log}
 \label{Observations}
 
\begin{tabular}{lcccccc}
\hline
Object & Dataset & Slit Width & Grating & PA & Spectral Res. & Spatial Res. \\
       &         & [arcsec]     &         & [$^{\circ}$]     & [km~s$^{-1}$] & [pc~pix$^{-1}$] \\
\hline

Mrk~3    & O5KS01010 & 0.1 & G430L & $-108$ & 250    & 13 \\
''       & O5KS01020 & 0.1 & G750L & $-108$ & 350    & '' \\
Mrk~348  & O5G405020 & 0.2 & G430L & 146  & 270    & 15 \\
Mrk~573  & O6BU02050 & 0.2 & G430L & $-71$  & 270    & 17 \\
''       & O6BU02040 & ''  & ''    & ''   & ''     & '' \\
''       & O6BU02020 & 0.2 & G750M & $-71$  & 40     & '' \\
''       & O6BU02010 & ''  & ''    & ''   & ''     & '' \\
''       & O6BU02030 & ''  & ''    & ''   & ''     & '' \\
NGC~1068 & O4WK010A0 & 0.1 & G430L & 202  & 250    & 4  \\
''       & O4WK010B0 & 0.1 & G750L & 202  & 350    & '' \\
NGC~3081 & O6BU06040 & 0.2 & G430L & $-110$ & 270    & 8  \\
''       & O6BU06050 & ''  & ''    & ''   & ''     & '' \\
''       & O6BU06030 & 0.2 & G750M & $-110$ & 40     & '' \\
''       & O6BU06010 & ''  & ''    & ''   & ''     & '' \\
''       & O6BU06020 & ''  & ''    & ''   & ''     & '' \\
NGC~3227 & O5KP01020 & 0.2 & G430L & $-150$ & 270    & 4  \\
''       & O5KP01040 & 0.2 & G750L & $-150$ & 370    & '' \\
NGC~4151 & O42302070 & 0.1 & G430L & 221  & 250    & 3  \\
''       & O42303050 & ''  & ''    & ''   & ''     & '' \\
''       & O42302080 & 0.1 & G750L & 221  & 350    & '' \\
''       & O42303060 & ''  & ''    & ''   & ''     & '' \\
''	 & O42305050 & 0.1 & G430L & 70   & 250    & 3  \\
''       & O42304090 & ''  & ''    & ''   & ''     & '' \\
''       & O42305060 & 0.1 & G750L & 70   & 350    & '' \\
''       & O423040a0 & ''  & ''    & ''   & ''     & '' \\
NGC~4507 & O5DF03010 & 0.2 & G430L & $-34$  & 270    & 12 \\
''       & O6BU17020 & 0.2 & G750M & $-138$ & 40     & '' \\
''       & O6BU17010 & ''  & ''    & ''   & ''     & '' \\
NGC~5643 & O6BU11040 & 0.2 & G430L & $-127$ & 270    & 4  \\
''       & O6BU11050 & ''  & ''    & ''   & ''     & '' \\
''       & O6BU11010 & 0.2 & G750M & $-127$ & 40     & '' \\
''       & O6BU11020 & ''  & ''    & ''   & ''     & '' \\
''       & O6BU11030 & ''  & ''    & ''   & ''     & ''\\
NGC~7682 & O6BU14040 & 0.2 & G430L & 18   & 270    & 15 \\
''       & O6BU14050 & ''  & ''    & ''   & ''     & ''\\
''       & O6BU14010 & 0.2 & G750M & 18   & 40     & ''\\    
''       & O6BU14020 & ''  & ''    & ''   & ''     & ''\\
''       & O6BU14030 & ''  & ''    & ''   & ''     & ''\\       
\hline

\end{tabular}
\end{minipage}
\end{table*}

One-dimensional (1D) spectra along the spatial direction were extracted from the 
2D frames for each object using the {\sc iraf} task apall. The nuclear spectra, 
centred at the peak of the continuum flux, were obtained by summing up the signal in a window size of 0.204\arcsec\ (4 pixels). In addition, several extractions in an aperture window 0.102\arcsec (2 pixels) long were done consecutively at both sides of the nucleus in the spatial direction. 
The number of extractions varied from object to object depending on the extent of the coronal 
line emission and other lines of the NLR.

The flux, full width at half maximum (FWHM), and peak position of the coronal lines [Ne\,{\sc v}]\l3425, [Fe\,{\sc vii}]\l3586, [Fe\,{\sc vii}]\l3760, [Fe\,{\sc xiv}]\l5303, [Fe\,{\sc vii}]\l6086, [Fe\,{\sc x}]\l6374, [S\,{\sc xii}]\l7611, and [Fe\,{\sc xi}]\l7892 were measured in each spectrum, if detected in the first place. If the region containing [Fe\,{\sc vii}]\l6086 was available for a given galaxy, measurements on that line were preferred over [Fe\,{\sc vii}]\l3586 and \FeVII\l3760 because the former 
is significantly stronger than the latter two. For comparison purposes, the low- and mid-ionisation lines [O\,{\sc ii}]\l3727, [Ne\,{\sc iii}]\l3869, and [O\,{\sc iii}]\l5007 were also measured.
In order to make an accurate subtraction of the continuum, we have selected continuum regions
approximately 300--600\kms\ away from the lines under consideration. For most targets we found that a straight line approximated the AGN continuum well, while in some sources (those with
the highest signal-to-noise ratios) the continuum was fit with a power law.
After subtraction of the interpolated continuum, it was assumed 
that the observed profile is well represented by a single or a sum of Gaussian components. 
The {\sc Liner} routine (Pogge \& Owen 1993) was used in this process. 
Overall, this approach was adequate for all spectra. 
Note that the spectra presented here are convolved with the STIS point-spread function (PSF). Although the STIS PSF for the CCD modes is narrow (${\rm FWHM}=2.3$ pixels at 7750~\AA), some contamination from the bright nuclear source to the extended regions is possible (see Nelson et al. 2000 for a detailed discussion of NGC~4151). We have analysed the PSF profile along the cross-dispersion axis, as traced by the point-like continuum emission of NGC~4151, and compared it to the spatial distribution of the different lines measured in the spectra presented here. We found that the PSF shape is much steeper than the slowly declining profile of the lines measured in the Seyfert~2 galaxies and therefore, contamination by the central brightest regions to the extended emission is negligible. Cecil et al. (2002) and Kraemer \& Crenshaw (2000a) reached a similar conclusion in the case of NGC~1068. 
However, the particular case of NGC~4151 is more complicated due to its very bright nucleus. While emission in the wings of the PSF contributes to the measurements next to the nucleus (up to 70\%), further out this contribution is of only a few percent (less than 30\%) and does not affect significantly our results.
Throughout this paper, we report the directly measured FWHMs of emission lines, rather than performing any instrumental correction, because the actual corrections depend significantly on the spatial extent of the emission region, and are different for point sources and extended sources (STIS Instrument Handbook$^1$);
and because at any given location, emission from extended and point-like regions might contribute to the emission-line profile.
In Table~\ref{Observations} we report the spectral resolution under the assumption of point-like emission. These values were calculated taking into account the spectral resolution (FWHM in pixels) of the 0.1$\arcsec$ and 0.2\arcsec\ slits for point sources and the spectral resolution provided by the different gratings. For extended sources, the instrumental FWHM can be of the order of 2--3 times greater than for point-like sources.

The velocity field along the slit, FWHMs and flux distributions of the lines of interest in the spatial direction were constructed for the objects of the sample. Because of the non-uniformity of the data, mostly when combining data gathered with the gratings G430L and G750M, only flux and centroid positions for the lines covered by the latter grating were employed.
The measured flux uncertainties are typically about 10\%, with some degradation in fainter regions ($<30$\%). However, as discussed above, at some distances from the nucleus of NGC~4151, the reported values are overestimated due to PSF effects, and we therefore consider a typical upper limit of 30\% for the overestimation of any reported line fluxes. In general, line centroids and FWHMs have errors of less than 150~km~s$^{-1}$. These errors were estimated using the standard deviation of five different measurements of the lines made with different continuum selections. However, the errors in the FWHM resulting from uncertainties in the spatial extent of the emission region (mentioned above) are much larger than this. We nevertheless report   FWHMs, because one can still identify systematic trends in FWHM changes, identify local abrupt changes, and compare FWHMs of lines of different degree of ionization.

\section{Results}\label{results}

In this section we will describe the general properties of the galaxies in the sample and the main results obtained from the measurements of the line fluxes for each galaxy. Emphasis is on the extent of the CLR and how its extent compares with gas of lower ionisation. To this purpose, we have plotted the flux distributions of the high-ionisation lines measured for the galaxies Mrk~573 (Fig.~\ref{mrk573_flux}), NGC~1068 (Fig.~\ref{ngc1068_flux}), Mrk~3 (Fig.~\ref{mrk3_flux}), NGC~4151 (Fig.~\ref{ngc4151_flux} and \ref{ngc4151_flux_70}), NGC~4507 (Fig.~\ref{ngc4507_flux}), and NGC~3081 (Fig.~\ref{ngc3081_flux}). We also added to the plots the flux distribution of the low-ionisation lines [Ne\,{\sc iii}], \OIII, and [O\,{\sc ii}] for comparison.
Additionally, Table~\ref{otras_fluxes} shows the line fluxes measured at different distances from the nucleus for the remaining galaxies in the sample, NGC~3227, Mrk~348, NGC~5643, and NGC~7682. Table~\ref{extension} shows the spatial extent to which each coronal line is detected as well as the total extent of the \OIII\ line emission along the slit. 
Notice that, unless stated otherwise, these extents correspond to the last position where the S/N was high enough for the line of interest to be detected (at $3 \sigma$ level). The possible presence of line emission further out from the nucleus was checked in two ways: first, we extracted additional spectra by summing up the signal from several pixels (up to 10) adjacent to the last spectrum where the line was detected. In all cases we found no emission further out.
Second, we used the upper limit of the fluxes of the lines of interest (derived from the continuum rms at the position where these lines should be located and an assumed FWHM) to compute line ratios involving the coronal line and a lower ionisation line present in the spectrum (e.g. \NeV/\OIII, \FeVII/\OIII, \FeX/[O\,{\sc i}]), and compared them with those measured closer to the nucleus.
In most cases, these upper limits are of the order or slightly lower than the line ratios measured closer to the nucleus. Therefore, although we can not discard emission further out, if there is any, it is very weak.
Also notice that the distances reported in this work can be affected by projection effects, for which we did not correct. For these reasons, the extents given in Table~\ref{extension} should be taken as  lower limits.

\subsection{Mrk~573}\label{mrk573_results}

This barred Seyfert 2 galaxy displays a very rich coronal line
spectrum, from optical to near-infrared wavelengths (e.g., Koski 1978; Tsvetanov \& Walsh 1992; Storchi-Bergmann et al. 1996; Riffel, Rodr\'iguez-Ardila \& Pastoriza 2006; Mullaney \& Ward 2008).
Recently, Ramos Almeida et al.
(2008) proposed that Mrk~573 is, in fact, a narrow-line Seyfert~1
galaxy (NLS1) rather than a Seyfert~2 based on the observations of O\,{\sc i} and 
Fe\,{\sc ii} permitted lines in the near-infrared. These features are characteristics
of only type~1 galaxies, as they are produced in the BLR. 
Nagao et al. (2004) reported spectropolarimetric observations of the NLR of this
source and found that it shows not only scattered broad H$\alpha$ emission
but also various narrow forbidden emission lines. The degree of polarisation
of the latter features is positively correlated with the ionisation potential
of the corresponding ion and the critical density of the corresponding
transition. They attributed these correlations to obscuration of the
stratified NLR by the geometrically and optically thick dusty torus.

The HST/STIS spectra of this galaxy were obtained with the slit oriented at a PA of 
$-71\deg$, that is, nearly aligned to both the [O\,{\sc iii}] ionisation cone (PA=120$\deg$)
reported by Regan \& Mulchaey (1999) and to the 
radio jet, whose NW component lies at PA=$-51\deg$ (Nagar et al. 1999). Previous
reports pointed out that the [O\,{\sc iii}] 
emission is extended along the same direction as the radio
emission (Haniff, Wilson \& Ward 1988; Pogge \& De Robertis 1995). The blue nuclear spectrum reveals strong coronal 
lines of [Ne\,{\sc v}], [Fe\,{\sc vii}], and [Fe\,{\sc xiv}], in addition to forbidden 
low- and medium-ionisation lines of [O\,{\sc ii}], [Ne\,{\sc iii}], 
and [O\,{\sc iii}], as well as recombination lines of H\,{\sc i} and He\,{\sc ii}. 
The red spectra display [Fe\,{\sc x}], along with low-ionisation
lines of [O\,{\sc i}], [N\,{\sc ii}] and [S\,{\sc ii}], as well as H$\alpha$. 
Previous studies using HST data (Ferruit et al. 1999;
Quillen et al. 1999; Schlesinger et al. 2009) have shown that the low- to medium-ionisation 
gas is extended to distances of up to $\sim$1100~pc to the SE and $\sim$1500~pc to the NW. In
addition, the circumnuclear region is rich in bright arcs and knots of
emission-line gas strongly aligned and interacting with a kiloparsec-scale low-power
radio outflow (Pogge \& De Robertis 1993; Falcke, Wilson \& Simpson 1998; Ferruit et al. 1999). 
Because all coronal lines we detected in this object are restricted to the inner 200~pc, 
that is, strictly at smaller scales than the region where the arcs and bright knots are located, we concentrate our attention only on the coronal features and discard any association
between the arcs and knots and the high-ionisation lines.

\begin{figure}
\includegraphics[width=\columnwidth]{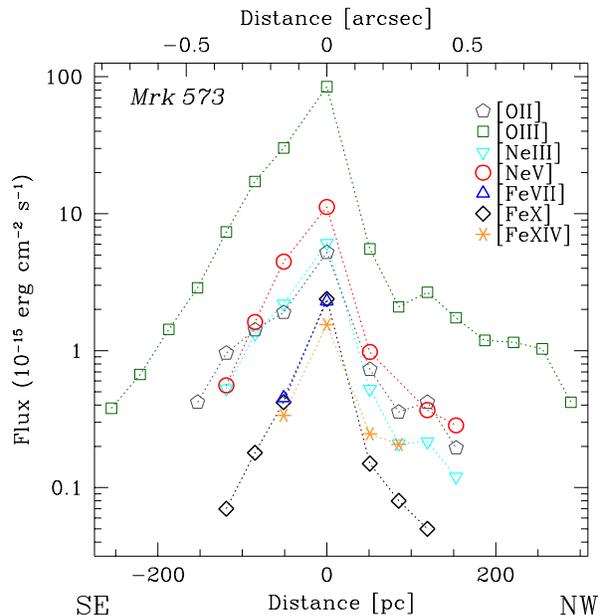}
\caption{Flux of the principal emission lines measured for Mrk~573 as a function of the projected distance to the nucleus. Typical error bars have the size of the symbols (see Section~\ref{Observations}) and are therefore not shown.}
\label{mrk573_flux}
\end{figure}

\begin{figure}
\includegraphics[width=\columnwidth]{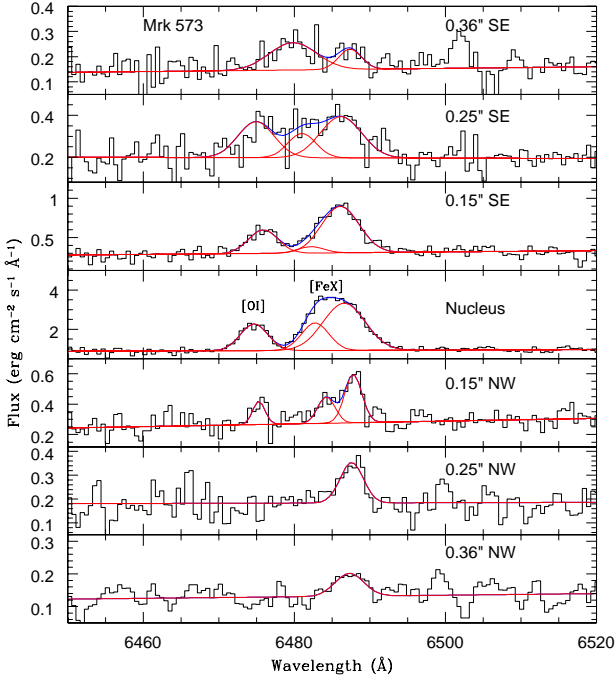}
\caption{[O\,{\sc i}] and \FeX\ profile fitting for the nucleus and adjacent regions of Mrk\,573.}
\label{mrk573_fe10}
\end{figure}

A close look at Fig.~\ref{mrk573_flux} shows that the CLR tends to be slightly more
extended towards the NW than to the SE. 
Moreover, the \OIII\ emission displays two small bumps at $\sim$120~pc and $\sim$250~pc NW, not
observed in the high-ionisation gas. To the SE, the extended emission drops more
smoothly. \NeV\ is the most extended CL, observed out to $\sim$160~pc NW 
and 120~pc SE from the nucleus (see Table~\ref{extension}). \FeVII~$\lambda$3586, 
in contrast, is detected only up to 50~pc SE. To the NW, it is restricted to the unresolved 
nucleus. However, it should be kept in mind that the true extent of the [Fe\,{\sc vii}] 
gas may be larger. Unfortunately, the only [Fe\,{\sc vii}] line covered by the spectra is 
in the blue region, where it is only a small fraction of [Fe\,{\sc vii}]\l6086, the 
strongest Fe$^{+6}$ line in the optical. For this reason, the true spatial extent of 
[Fe\,{\sc vii}] is inconclusive. In fact, Storchi-Bergmann et al. (1996) presented high signal-to-noise spectra of Mrk~573 and reported the detection of \FeVII~$\lambda$6086 and \NeV\l3425 at a distance of 6\arcsec\ NW and SE from the nucleus. \FeVII~$\lambda$3586 was not reported.

The blue spectra also include [Fe\,{\sc xiv}]\l5303, the highest 
ionisation line in the optical region detected in Mrk~573. It is observed in the unresolved
nuclear spectrum and its emission extends $\sim$80~pc to the NW and $\sim$50~pc 
to the SE. 

The red spectrum, which has a better spectral resolution than the blue ones, 
contains [Fe\,{\sc x}]~$\lambda$6374. This line is prominent in the 
nucleus, displaying a conspicuous broad flat top profile
(see Fig.~\ref{mrk573_fe10}) typical of a double-peaked structure, in
contrast to the Gaussian-like profile seen in [O\,{\sc i}] and [S\,{\sc ii}].
The double-peaked nature of [Fe\,{\sc x}], firstly recognised by
Schlesinger et al. (2009) in the nuclear spectrum of Mrk~573, 
became evident when we examined the line
profile in the different apertures, shown
in Fig.~\ref{mrk573_fe10}. It can be seen that the blue
peak becomes progressively more prominent than the red one from the NW to the SE
except at 0.15$\arcsec$SE, where its intensity, relative to the red peak, decreases.
Moreover, the relative separation between both peaks increases: while in the
nuclear aperture they are separated by $\sim$180~km\,s$^{-1}$, at 0.36$\arcsec$ SE,
the last position where [Fe\,{\sc x}] is detected, they are separated by $\sim$350~km\,s$^{-1}$.
From Fig.~\ref{mrk573_fe10} it is also easy to see that [Fe\,{\sc x}] extends up to 
120~pc both NW and SE. For comparison, [O\,{\sc i}] is detected up to 
150~pc SE and 190~pc NW (these apertures are not shown in the plot). However, there
are positions (0.36$\arcsec$NW, 0.25$\arcsec$NW, and 0.36$\arcsec$SE) where
no [O\,{\sc i}] emission is seen. Moreover, the profiles of the other lines seem  
rather different from that of [Fe\,{\sc x}], leading us to suggest that this
emission should be related to a high-ionisation matted-bounded cloud that has been radially accelerated.
It is worth to mention that at 85~pc SE, the line profiles of H$\alpha$, [N\,{\sc ii}], and [S\,{\sc ii}] are very complex, with multiple components in each line 
while that of [Fe\,{\sc x}] is dominated by the red peak. 

The increase in relative separation between the red and blue peaks of [Fe\,{\sc x}] with the distance when going from NW to SE is unique. To the best of our knowledge no previous report of such a behaviour in an optical emission line with IP$>200$~eV is found in the literature. 
Moreover, the detection of [Fe\,{\sc x}] to distances of up to 120~pc, which we report here, is remarkable, and puts tight constraints on photoionisation models. This issue will be examined in more detail in Section~\ref{photo}.

\subsection{NGC~1068}\label{ngc1068_results}

\begin{figure}
\includegraphics[width=\columnwidth]{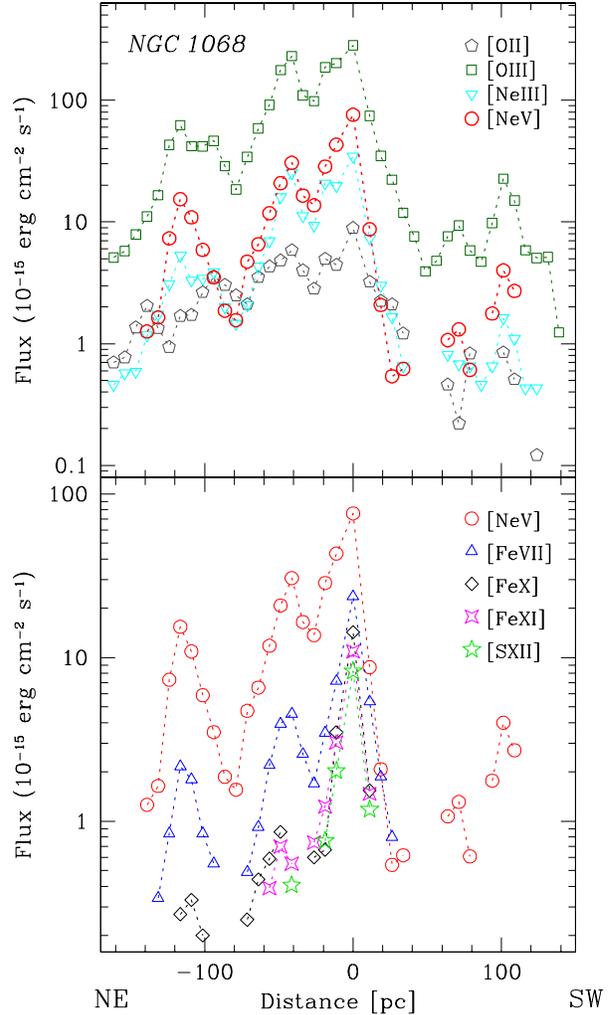}
\caption{Flux of the principal emission lines of NGC~1068 as a function of the projected  distance to the nucleus. For visualisation purposes we plotted the line flux distributions in two panels: in the upper panel we show \NeV\ and low-ionisation lines and in the lower panel \NeV\ and high-ionisation lines. Typical error bars have the size of the symbols (see Section~\ref{Observations}).}
\label{ngc1068_flux}
\end{figure}

NGC~1068 is one of the nearest Seyfert galaxies and is usually considered as the 
archetypal Seyfert 2 galaxy. It has been extensively studied in every wavelength range, 
from X-rays to the radio regime (e.g., Oliva \& Moorwood 1990; Miller, Goodrich \& Mathews 1991; Antonucci, Hurt \& Miller 1994; Machetto et al. 1994; Gallimore et al. 1996; Young et al. 1995; Marconi et al. 1996; Young, Wilson \& Shopbell 2001; Kinkhabwala et al. 2002; Jaffe et al. 2004). The structure of the inner regions of this galaxy is extremely complex. It is classified as a (R)SA(rs)b galaxy, showing a strong and large inner bar inside a very large, weak outer bar/oval disk. A cone-like NLR oriented roughly in the northeast-southwest direction was resolved into filaments and point sources by HST narrow band imaging (Evans et al. 1991; 
Macchetto et al. 1994). Spatially coincident with the NLR, VLA maps revealed a 13\arcsec\ radio bipolar structure constituted by jets ending in radio lobes along the PA = 30$^\circ$ (Wilson \& 
Ulvestad 1983). On sub-arcsecond scales, the jet is resolved into four components 
(Gallimore et al. 1996). Two of these components are located roughly in the north-south direction. 
The component located more to the north is believed to be the hidden active nucleus. At a 
distance of $\sim$ 0.3\arcsec\ from it, towards the north (along the PA = 10$^\circ$), the third 
component is located. Here, the jet seems to have an abrupt change in direction and the fourth  
component is observed at a distance of $\sim$0.3\arcsec\ from the third along the PA = 30$^\circ$.
NGC~1068 is also a strong source of X-ray emission. Chandra high-resolution images showed an  
impressive agreement between the X-ray emission and the [O\,{\sc iii}] images on small and 
large scales, with a detailed correspondence between both (Young et al. 2001). These images 
showed a bright nucleus, extended bright emission toward the NE, and large scale structure 
reaching at least 1 arcmin to the NE and SW. 

The complexity of this galaxy is also evident in the line profiles and fluxes, showing 
strong variations between adjacent spectra along the slit (see Fig.~\ref{ngc1068_flux}). Notice that the STIS slit was located along the ionisation cone and roughly coincident with the nuclear radio jet (see Table~\ref{Observations}). The STIS spectra show several coronal 
lines in the 2900--10300~\AA\ range, from [Ne\,{\sc v}] and [Fe\,{\sc vii}] lines, whose 
ionisation potential are $\sim$ 100~eV, to the extreme ionisation [S\,{\sc xii}]\l7611 line, 
with an ionisation potential of 504.7~eV. This very high-ionisation line has been reported only in NGC~1068 (Kraemer \& Crenshaw 2000b).

The most extended CLs, \NeV\l3425 and \FeVII\l6086, exhibit similar 
spatial morphology as the intense \OIII\l5007 line, showing several blobs 
located mainly in the NE side of the nucleus. The \NeV\ line extends up to 140~pc NE, with two peaks of emission located at 40~pc and 115~pc NE of the 
nucleus. Towards the SW this line is relatively weak and vanishes and re-appears several 
times. The maximum extent towards the SW is 110~pc. The \FeVII\l6086 line is 
observed up to 130~pc NE of the nucleus, though it is not detected from 80~pc to 85~pc. 
At the SW side, this line is much less extended, observed only up to 25~pc.
At $\sim$20~pc NE enhanced emission is observed in the low-ionisation lines plotted in Fig.~\ref{ngc1068_flux}. This blob is not followed by the high-ionisation lines. 

The \FeX\l6374 line is more complicated to analyse due to the strong blend with the 
[O\,{\sc i}]\l6363 line. However, fixing the parameters (FWHM, centroid position, and intensity) of the latter, as determined by the near [O\,{\sc i}]\l6300 line, which belong to the same doublet, allowed us to measure the size of the \FeX\ emission region. Towards the SW this line is observed only next to the nucleus, at 10~pc. On the other hand, the extent towards the NE is quiet large, extending almost as much as the \FeVII\ line. It is clearly present next to the 
nucleus (10~pc NE), from 50~pc to 65~pc, and from 110~pc to 115~pc.
The higher ionisation regions, traced by \FeXI\l7892 and 
[S\,{\sc xii}]\l7611 lines, are much more compact but still spatially resolved. These lines 
extend up to $\sim$10~pc to the SW, and towards the NE up to 50~pc and 20~pc, 
respectively. In general, when the high-ionisation lines are observed, they tend to follow the same flux distribution as the lower ionisation lines (i.e. \OIII).

It is interesting to note that, towards the NE, the size of the extended emission decreases as the ionisation potential of the ``coronal ions'' increases, in complete agreement with photoionisation by a central source. However, towards the SW the lines show an entirely different behaviour that cannot be explained simply by anisotropy of the ionising radiation. Although \NeV\l3425 and \FeVII\l6086 have almost the same ionisation potential and critical densities, the first one extends much more towards the SW than the second one. 
Similar results have been found in the NIR. Geballe et al. (2009) reported $L-$ and $M-$band spectroscopy of the nucleus of NGC~1068 obtained at 0.3$\arcsec$ (20~pc) resolution with the spectrograph slit aligned approximately along the ionisation cones of the AGN. They found that all CLs detected in that wavelength region peak 20~pc north of the AGN and are very weak or not present towards the south. 
This can be understood in terms of the strong extinction towards the SW measured in this object. Assuming a case B recombination ratio of $\rm{H}\alpha/\rm{H}\beta = 3.1$ (Osterbrock 1989) we determined the reddening as a function of the distance to the nucleus. Towards the NE the measured ratio is close to the theoretical ratio, therefore reddening is not strongly affecting the gas. However, towards the SW the reddening increases with the distance up to {\it E(V-B)}$=0.9$ measured at 35~pc SW. Further out, no determination of the reddening was possible due to the weakness of the hydrogen lines.
Assuming that in the absence of reddening the values of the ratio \NeV\l3425/\FeVII\l6086 towards the SW are similar to the ones measured to the NE,
\NeV\l3425 is, intrinsically, at least 10 times stronger than \FeVII\l6086. Clearly, the 
dusty environment to the SW extinguishes completely the latter line. 

The \FeVII\ 3759/6086 and 5159/6086~\AA\ line ratios can be use as diagnostics of the density in the Fe$^{+6}$ emitting region (Keenan \& Norrington 1987). Of all the galaxies in the sample, NGC~1068 is the only one for which these lines are available. However, due to the weakness of the lines in the blue spectra and the uncertainties due to reddening corrections (specially in the case of the former ratio which is strongly sensitive to extinction), we were not able to derive a reliable value for the density of the coronal-line gas.

\subsection{Mrk~3}\label{mrk3_results}

Mrk~3 is a nearby S0 galaxy. Although it is classified as a Seyfert 2 galaxy from optical spectroscopy, it shows evidence of a hidden Seyfert 1 nucleus in polarised light (Miller \& Goodrich 1990). The morphology of the inner regions of Mrk~3 is highly intricate. Narrow band 
HST imaging, centred at [O\,{\sc ii}]\l3727, [O\,{\sc iii}]\l5007, H$\gamma$, and H$\alpha$, 
showed an S-shaped NLR extending across the inner 2$\arcsec$ in the east-west direction 
(PA$= 70^\circ$ east of north) and composed of a large number of resolved 
knots (Capetti et al. 1995). This small scale structure is embedded in a more 
extended ($10\arcsec$) biconical emission region aligned along PA=114$^\circ$ 
(Pogge \& De Robertis 1993). Moreover, high resolution radio observations 
revealed two highly collimated radio jets lying along PA$=84^\circ$ (Kukula et al. 1993). The jet shows a slight S-shape curvature expanding about 2$\arcsec$, the west side ends in an extended lobe containing a hot spot at 1.2$\arcsec$ from the nucleus. 
The soft X-ray emission of Mrk~3 is also spatially elongated along the 
[O\,{\sc iii}] emission (Sako et al. 2000).

\begin{figure}
\includegraphics[width=\columnwidth]{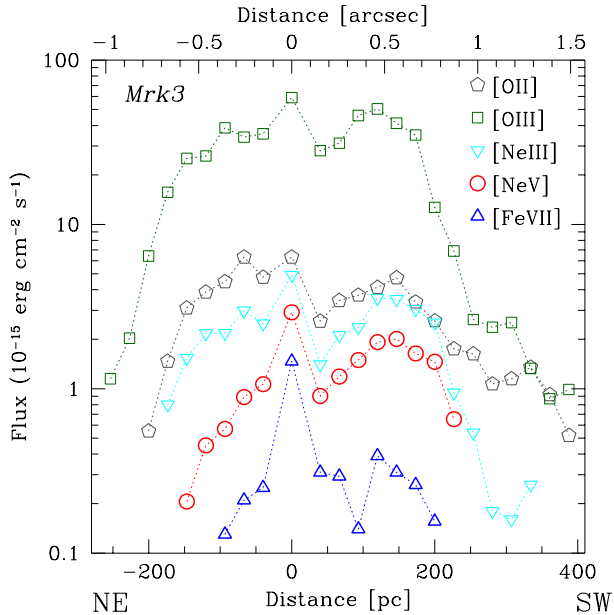}
\caption{Same as Fig.~\ref{mrk573_flux}, but for Mrk~3.}
\label{mrk3_flux}
\end{figure}

Two different gratings were used in the STIS observations of Mrk~3 (G430L and G750L), 
with the slit following the S-shaped NLR at PA=73$^\circ$ (see Table~\ref{Observations}) 
and nearly aligned with the radio jet. An inspection of the optical spectrum of this galaxy 
shows that it displays very weak coronal line emission. Indeed, in the blue 
region, only [Ne\,{\sc v}]$\lambda\lambda$3343,3425 and faint 
[Fe\,{\sc vii}]\l5158 are observed. In the red spectra, [Fe\,{\sc vii}]\l5722 
and \l6086 are detected. Higher ionisation lines are not present in the optical spectrum of Mrk~3 and, to our knowledge, neither in the infrared region (Osterbrock, Shaw \& Veilleux 1990; Wilson \& Nath 1990; Heisler \& De Robertis 1999; Knop et al. 2001). This makes Mrk~3 one of the lowest ionisation galaxies of the whole sample. Interestingly, Mrk~3 exhibits the most extended CLR of the ten AGN studied here (see Table~\ref{extension}). The [Ne\,{\sc v}]\l3425 emission, for instance, extends up to 230~pc to the SW from the nucleus and 150~pc to the NE. Moreover, [Fe\,{\sc vii}]\l6086 can be traced up to 200~pc to the SW and 90~pc to the NE. 
In contrast, medium ionisation lines such as [O\,{\sc iii}]\l5007 
extend up to 650~pc SW and 450~pc NE.

Similar to other lines, [Ne\,{\sc v}]\l3425 and [Fe\,{\sc vii}]\l6086  
are stronger towards the SW, both displaying a second peak of emission which does not coincide spatially: 150~pc from the centre for the former and 120~pc for the latter.
The flux distribution for the most conspicuous lines, including the coronal lines
plotted in Fig.~\ref{mrk3_flux}, shows a similar pattern for the five species studied. 

The lack of highly ionised species in the optical spectrum contrasts to what is
observed at high energies. The observed soft X-ray spectrum of Mrk~3, for instance, is dominated
by line emission from highly ionised stages of the lighter metals (Pounds \& Page 2005).
These authors stress the similarity between the X-ray line spectrum of this
object and that of NGC~1068 but recall that the fluxes are an order of magnitude
less, mainly due to Mrk~3 lying at a greater distance and the attenuation at
the longer wavelengths by the relatively large Galactic column in the direction of
this source. As we already saw in Section~\ref{ngc1068_results},
the optical spectrum of NGC~1068 displays conspicuous high-ionisation
lines, not seen in Mrk~3. This result leads us to speculate if the lack of
high-ionisation lines in the optical spectrum of Mrk~3 is due to technical reasons
(not enough S/N) rather than to physical reasons. In order to test this scenario
we assume that, as in the X-ray region, the optical spectrum of Mrk~3 is very similar 
to that of NGC~1068. Then, line ratios among coronal lines should be very similar 
in both objects, allowing us to determine an upper limit to the flux of \FeX\ in Mrk~3.
The nuclear ratio [Fe\,{\sc x}]/[Fe\,{\sc vii}] measured in NGC~1068 is 0.60$\pm$0.08. Assuming a similar ratio in Mrk~3 and a measured line flux of 8.53$\pm$1.40$\times10^{-16}$ erg\,cm$^{-2}$\,s$^{-1}$ for [Fe\,{\sc vii}], the expected upper limit
for the [Fe\,{\sc x}] flux is 5.1$\times10^{-16}$ erg\,cm$^{-2}$\,s$^{-1}$. This flux is nearly six times the RMS (3$\sigma$) of the observed adjacent continuum so, if present, this line should have been detected. Therefore, we conclude that the lack of high-ionisation 
lines in Mrk~3 is intrinsic, probably due to physical reasons rather than to the low S/N.

\subsection{NGC~4151}\label{ngc4151_results}

NGC~4151 is a SABab galaxy that harbours a type 1 Seyfert nucleus. 
It is one of best studied AGNs in the literature at all wavelengths intervals and in almost every aspect of its observed spectrum (e.g., Schulz 1995; Ulrich 2000, and references therein). Due to its proximity, the NLR can be resolved from the ground, even with telescopes without AO. Ulrich (1973) showed that the NLR extends to several arcsecs and consists of at least four distinct clouds. Pogge (1989) and P\'erez et al. (1989) detected [O\,{\sc iii}] emission in ground-based images, finding it to be extended by $\sim20\arcsec$ along PA 228$\deg$. The [O\,{\sc iii}] emission of this galaxy has a biconical morphology (Schulz 1988) along
PA 55$\deg$, clearly not aligned with the radio axis at PA $\sim$ 77$\deg$ 
(Ulvestad, Wilson \& Sramek 1981). The total 
size of the radio emission is $\sim$5.5$\arcsec$ (350~pc), 3$\arcsec$ along the
southwest direction and 2.5$\arcsec$ along the northeast. The total opening angle of
the cones is 70$\deg$. Coronal lines in this object have been reported from the X-ray to optical region (e.g., Penston et al. 1984; Sturm et al. 2002; Ogle et al. 2000)
but, to our knowledge, no studies have shown in detail the spatial extent and kinematics 
of the high-ionisation gas at scales $<100$~pc, except for the recent work presented by Storchi-Bergmann et al. (2009). These authors, using AO Gemini/NIFS 3D spectroscopy, mapped NIR coronal lines, showing that the emission region of these lines were no larger than 50~pc.

The HST/STIS spectra of NGC~4151 were taken at two position angles. One position, at  
PA=221$\deg$, intersecting the emission cone and passing through the
nucleus. The other position, PA=70$\deg$, was offset by 0.1\arcsec\ to the south of the nucleus and coincides with the radio jet observed in this galaxy. 
Because of this configuration, it allows us to fully explore the extent of the CLR and compare its size with that of other lines emitted by the NLR, and to study the effects of the jet on the high-ionisation gas.

\begin{figure}
\includegraphics[width=\columnwidth]{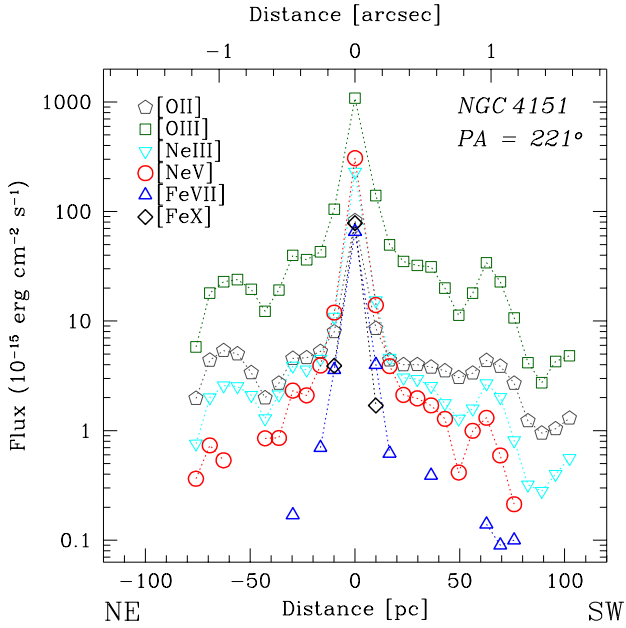}
\caption{Same as Fig.~\ref{mrk573_flux}, but for NGC~4151 at PA = 221$^{\circ}$.}
\label{ngc4151_flux}
\end{figure}

\begin{figure}
\includegraphics[width=\columnwidth]{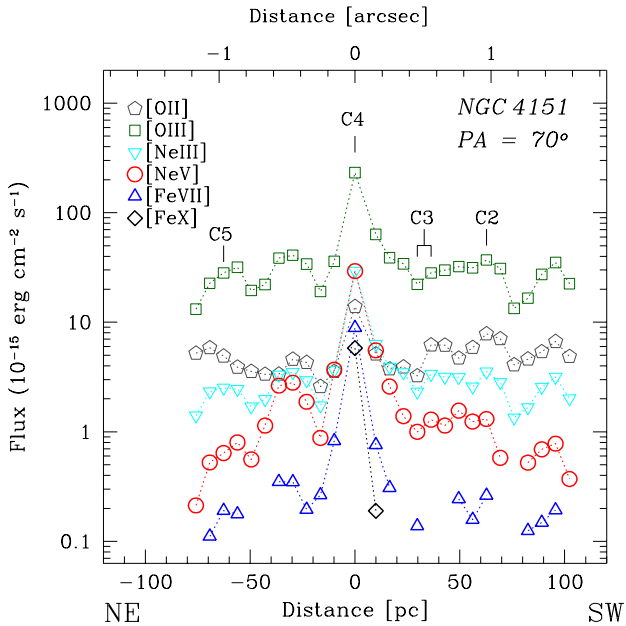}
\caption{Same as Fig.~\ref{mrk573_flux}, but for NGC~4151 at PA = 70$^{\circ}$. The labels C2 to C5 indicate the position of the radio knots intersected by the slit.}
\label{ngc4151_flux_70}
\end{figure}

Fig.~\ref{ngc4151_flux} and Fig.~\ref{ngc4151_flux_70} show the flux distribution of [O\,{\sc ii}], [O\,{\sc iii}], [Ne\,{\sc iii}], [Ne\,{\sc v}], [Fe\,{\sc vii}], and [Fe\,{\sc x}]
in the spatial direction with the slit aligned along PA=221$\deg$ and PA=70$\deg$, respectively. 
Overall, the flux distributions of the CLs resemble those of the lower ionisation lines. Several peaks of emission are observed at both sides of the nucleus, making evident the knotty structure of the NLR in this object. The fluxes measured along both position angles do not show any significant difference, although the fluxes for the inner $\sim 40$~pc along PA=221$\deg$ are slightly higher than the ones along PA=70$\deg$. This can be explained as a result of the offset in the position of the second slit. 

In terms of the size of the coronal emission region, we found that it is restricted to the inner $\sim 200$~pc. The \NeV\ line is the most extended. This line is detected up to 75~pc both SW and NE along PA=221$\deg$, and 75~pc NE and 100~pc SW along PA=$70\deg$. The region emitting [Fe\,{\sc vii}], in contrast, is asymmetric. Along PA=$221\deg$ it is detected up to 75~pc SW but only up to 30~pc NE, while along PA=$70\deg$ this line is more extended, reaching 70~pc NE and 95~pc SW.
[Fe\,{\sc x}] is much more compact, being limited to the inner radius of 10~pc from the centre along PA=221$\deg$ and to the nucleus and next to it towards the SW along PA=70$\deg$. 

\subsection{NGC~4507}\label{results_4507}

NGC~4507 is a barred spiral galaxy. It was originally classified as a type 2 Seyfert galaxy (Durret \& Bergeron 1986) but later re-classified as Seyfert 1.9 (Veron-Cetty \& Veron 1998) because of the presence of broad H$\alpha$ observed sometimes in its spectrum.
Several high-ionisation lines have previously been detected in  this 
galaxy. In the optical region [Ne\,{\sc v}], [Fe\,{\sc vii}], [Fe\,{\sc x}], 
[Fe\,{\sc xi}], and [Fe\,{\sc xiv}] were 
reported by Penston et al. (1984), Durret \& Bergeron (1986), and Appenzeller 
\& \"{O}streicher (1988). In the infrared, [Si\,{\sc vi}]~1.962~$\mu$m (Marconi et al. 1994), 
[Si\,{\sc ix}]~3.94~$\mu$m (Lutz et al. 2002), and [Ne\,{\sc v}]~14.3~$\mu$m (Gorjian et al. 2007) has been detected.

The most prominent lines observed in the STIS nuclear blue spectrum of NGC~4507 are 
[O\,{\sc iii}]\l5007 and [Ne\,{\sc iii}]\l3869, followed by the high-ionisation line 
[Ne\,{\sc v}]\l3425. Other coronal lines observed are [Fe\,{\sc vii}]\l3586, 
[Fe\,{\sc vii}]\l3759, [Fe\,{\sc vii}]\l5158, and [Fe\,{\sc xiv}]\l5303. 

\begin{figure}
\includegraphics[width=\columnwidth]{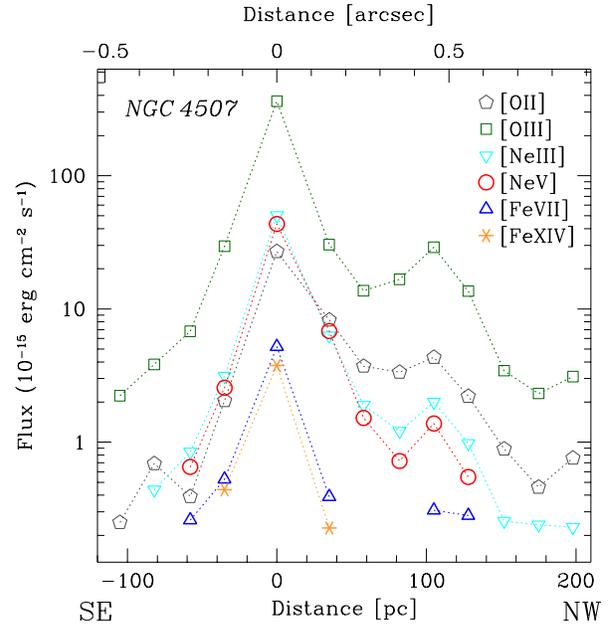}
\caption{Same as Fig.~\ref{mrk573_flux}, but for NGC~4507.}
\label{ngc4507_flux}
\end{figure}

Fig.~\ref{ngc4507_flux} shows the flux variation of the most conspicuous lines versus the distance from the centre. It can be seen that the emission is asymmetric, with a strong elongation towards NW and a second peak of emission at $\sim 110$~pc (0.46$\arcsec$) from the nucleus. This second peak agrees in position with the [O\,{\sc iii}] HST image from Schmitt et al. (2003), which shows elongated emission along PA=$-35^\circ$ in the inner $\sim$2$\arcsec$ 
region and a blob of emission located at 1$\arcsec$ to the NW. Notice that this PA is very close to the one used in the STIS observation of the blue spectra (see Table~\ref{Observations}). 
Inspection of the 2D STIS spectrum clearly shows this enhanced emission at 1$\arcsec$
NW ($\sim 230$~pc) from the nucleus for the strongest lines (i.e. \OIII, H$\beta$, 
[Ne\,{\sc iii}], [O\,{\sc ii}]). Unfortunately we are not able to tell whether or not \NeV\ is present at this distance due to the presence of a cosmic ray. 
\FeVII\ emission is not seen at this distance from the nucleus in the STIS spectrum. 
Even so, this galaxy shows one of the most extended \FeVII\ regions of the whole 
sample (as traced by the \FeVII\l3586 line), only exceeded by the \FeVII\l6086 
line displayed by Mrk~3 (see Table~\ref{extension}). The \FeVII\ region extends as much as the \NeV\l3425 region, reaching $\sim 130$~pc to the NW and $\sim 60$~pc to the SE. Towards the NW the \FeVII\ line disappears at $\sim 60$~pc and is observed again at $\sim 100$~pc.
On the other hand, the \FeXIV\ emission-line gas is located closer to the nucleus. 
It can only be traced up to $\sim 35$~pc at both sides of the nucleus. 

\begin{figure}
\includegraphics[width=\columnwidth]{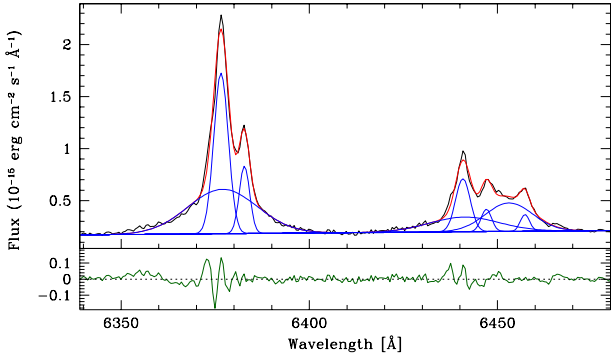}
\caption{[O\,{\sc i}]\l\l6300,6363 and \FeX\ profiles of NGC~4507 decomposed into gaussian components.}
\label{ajuste_fex}
\end{figure}

NGC~4507 was also observed at higher spectral resolution with the G750M grating. Notice however that the blue spectrum was taken with a different PA than the red spectra (see Table~\ref{Observations}). For this reason the \FeX\ line was not included in Fig.~\ref{ngc4507_flux}. The spectra show lines with very complex structures, displaying multiple components with strong variations along the slit.
The \FeX\ line is strongly blended with the [O\,{\sc i}]\l6363 line. In order to deblend these lines we applied constraints in the [O\,{\sc i}]\l6363 line fitting, assuming the same FWHM, velocity shift, and 1/3 of the flux of the stronger [O\,{\sc i}]\l6300 line. 
The nuclear spectrum shows a very intricate [O\,{\sc i}]$+$\FeX\ profile. The [O\,{\sc i}] lines were fitted by three Gaussian components, with velocity shifts (FWHMs) of 90\kms\ (200\kms), 110\kms\ (1000\kms), and 380\kms\ (150\kms). 
Two Gaussian components were required to obtain a good fit of the nuclear \FeX\ line: one broad component ($\rmn{FWHM}=685$\kms) redshifted by $\Delta\rmn{V}=155$\kms\ and a narrow component ($\rmn{FWHM}=127$\kms) redshifted by $\Delta\rmn{V}=352$\kms. 
The total flux in \FeX\ is $(4.8 \pm 0.3) \times 10^{-15}$ erg\,cm$^{-2}$\,s$^{-1}$. In Fig.~\ref{ajuste_fex} we show the profile fitting of the [O\,{\sc i}]-\FeX\ complex. NGC~4507 and Mrk~573 (see Section~\ref{mrk573_results}) are the only galaxies in the sample displaying such a particular \FeX\ profile.
\FeX\ is also observed at both sides of the nucleus, with fluxes of $(0.26 \pm 0.13) \times 10^{-15}$ erg\,cm$^{-2}$\,s$^{-1}$ at $\sim$ 35~pc NW and SE.

\subsection{NGC~3081}

\begin{figure}
\includegraphics[width=\columnwidth]{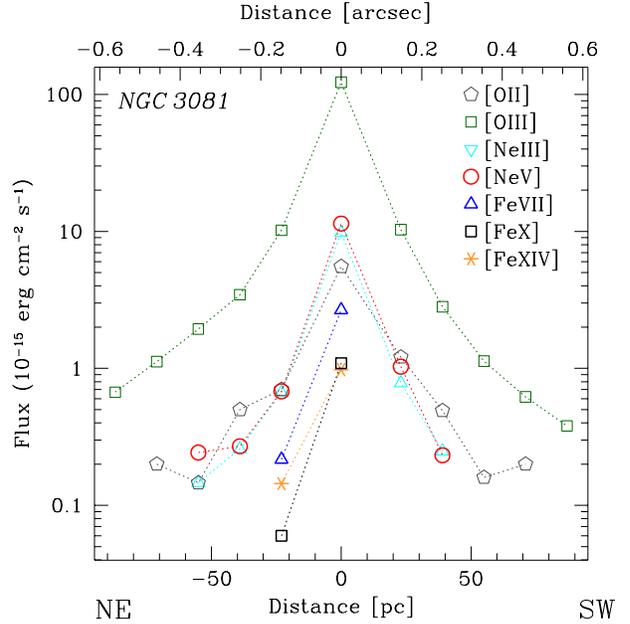}
\caption{Same as Fig.~\ref{mrk573_flux}, but for NGC~3081.}
\label{ngc3081_flux}
\end{figure}

NGC~3081 is a galaxy with complex morphology showing multiple stellar and dust rings, and a weak bar, hosting a Seyfert type 2 nucleus (Buta 1990). A spectropolarimetric study by Moran et al. (2000) has established that it is a Seyfert 1, obscured by dense material at a few
parsecs from the nucleus. The radio emission of this galaxy has been studied by Ulvestad \& Wilson (1989). VLA observations at 6~cm and 20~cm (flux densities of 0.9 and 2.5~mJy, respectively) revealed an unresolved source.

In the optical region, in addition to strong permitted and low-ionisation 
forbidden lines, Appenzeller \& \"{O}streicher (1988) reported the detection
of [Fe\,{\sc vii}], [Fe\,{\sc x}], [Fe\,{\sc xi}], and [Fe\,{\sc xiv}], although
this latter feature is strongly blended with [Ca\,{\sc v}].  
In the infrared region, Lutz et al. (2002) detected [Si\,{\sc ix}]~3.94~\micron\ (IP = 303~eV). Reunanen, Kotilainen \& Prieto (2003) reported long-slit NIR spectroscopy on 
this object along the direction of the ionisation cone and perpendicular to it. They found strong nuclear coronal lines of [Si\,{\sc vi}]~1.964~$\mu$m and [Si\,{\sc vii}]~2.484~$\mu$m, as
well as the detection of [Ca\,{\sc viii}]~2.321~$\mu$m and [Al\,{\sc ix}]~2.043~$\mu$m. The coronal [Si\,{\sc vi}] line (IP = 166~eV) showed the most extended emission in the 1.5--2.5~\micron\ range studied by these authors, with a flux distribution FWHM of 350~pc and 200~pc parallel and perpendicular to the ionisation cone respectively, suggesting a strongly  anisotropic radiation field. 

The STIS blue spectra of NGC~3081 show strong, very symmetric, single-peaked lines from low-ionisation (i.e. [O\,{\sc ii}], [Ne\,{\sc iii}], and [O\,{\sc iii}]) and high-ionisation species (i.e \NeV, \FeVII, and \FeXIV), as well as permitted hydrogen and helium lines.
The most extended CL is \NeV, which can be traced up to $\sim$40~pc SW and $\sim$55~pc NE 
of the nucleus. It can be seen in Fig.~\ref{ngc3081_flux} that its flux distribution is nearly symmetrical with respect to the nucleus. 
The spatial distribution of the lower ionisation gas is also symmetric relative to the maximum of emission located at the unresolved nucleus and compact. \OIII\ displays the largest 
emission region, extending up to $\sim$230~pc SW and NE of the nucleus. The CLs of \FeVII\l3760 and \FeXIV\ are only detected at the nucleus and next to it towards the NE. No extended emission of these lines is detected to the SW.

The red spectra, of higher spectral resolution, show lines with more complex profiles at both directions from the nucleus. The nuclear spectrum displays a relatively weak, but clearly present, \FeX\ line. However, due to the weakness of this line, little can be said about its detailed profile. As in the case of the other coronal lines, the \FeX\ emitting region is also restricted to the inner region, detected only in the nucleus and at 25~pc NE from it.

It is worth mentioning that the position of the STIS slit is the same as in the observation  presented by Reunanen et al. (2003), that is, perpendicular to the cone. However, the coronal emission observed by these authors is much more extended than the one displayed by the optical coronal lines in the STIS data. This can be explained if we consider that the spatial resolution of the Reunanen et al. (2003) data for NGC~3081 is 1$\arcsec$ and that the slit width they employed is a factor of ten larger than that of STIS.

\subsection{NGC3227, Mrk~348, NGC~5643, and NGC~7682}

Although these galaxies exhibit signs of coronal line emission, because of the weakness of these lines or due to the poorer spatial resolution, the analysis of their coronal line emission is restricted to only a few apertures. 
In Table~\ref{otras_fluxes} we report the measurements of the most important lines observed in these galaxies. Below we describe the most important aspects of the HST/STIS spectra of these sources. 

\begin{table*}
\begin{minipage}{126mm}
\caption{Emission line fluxes for Mrk~348, NGC~5643, NGC~3227, and NGC~7682.}
\label{otras_fluxes}
\begin{tabular}{llcccccc}
\hline
Object     & Distance & [Ne\,{\sc v}] & [O\,{\sc ii}] & [Ne\,{\sc iii}] & [O\,{\sc iii}] & [Fe\,{\sc vii}] & [Fe\,{\sc x}] \\
       & [arcsec] & $\lambda 3425$ & $\lambda 3727$ &  $\lambda 3869$ & $\lambda 5007$ & $\lambda 6086$ & $\lambda 6374$\\
\hline
Mrk~348  & 0.0       & 4.66(0.32) & 18.9(0.34)    & 10.9(0.33)    & 148(0.62) & -- & \\

      & 0.15 NW & 2.14(0.25) & 7.61(0.20)    & 4.47(0.19)    & 49.5(0.16) & -- &  \\
      & 0.25 NW & 1.81(0.20) & 6.96(0.19)    & 3.69(0.14)    & 35.2(0.11) & -- &  \\
      & 0.36 NW & 0.74(0.19) & 3.38(0.17)    & 1.33(0.14)    & 7.45(0.10) & -- &  \\
\\
NGC~3227 & 0.00  & 20.00(4.10) & 11.90(3.50) & 26.70(4.33) & 222(2.50)  & 8.2(2.7) & 8.3(3.4)\\

\\
NGC~5643 & 0.0        & 1.14(0.18) & 3.49(0.21)    & 2.10(0.20)    & 43.3(0.09) & -- & 2.40(1.10)\\

      & 0.15 NE & 1.49(0.16) & 2.74(0.10)    & 1.43(0.09)    & 24.0(0.03) & -- & --\\
      & 0.25 NE & 1.28(0.15) & 2.30(0.15)    & 1.23(0.14)    & 14.2(0.09) & -- & --\\
      & 0.36 NE & 0.60(0.12) & 1.81(0.09)    & 0.62(0.08)    & 7.65(0.05) & -- & --\\
\\
NGC~7682 & 0.0        & 0.86(0.38) & 3.82(0.12)    & 2.70(0.13)    & 25.0(0.13) & -- & --\\
      & 0.15 NE  & 0.31(0.14) & 1.03(0.07)    & 0.47(0.06)    & 4.10(0.05) & -- & --\\
      & 0.25 NE  & --     & 1.04(0.05)    & 0.45(0.04)    & 4.57(0.04) & -- & --\\
      & 0.36 NE  & 0.26(0.06) & 2.27(0.07)    & 0.79(0.06)    & 7.78(0.04) & -- & --\\
      & 0.46 NE  & 0.25(0.09) & 2.11(0.05)    & 0.70(0.04)    & 7.13(0.04) & -- & --\\

      & 0.15 SW  & 0.17(0.01) & 0.92(0.06)    & 0.23(0.05)    & 4.43(0.10) & -- & --\\
\hline
\end{tabular}
\\
\medskip
NOTE -- Fluxes are given in units of $10^{-15}$~erg~cm$^{-2}$~s$^{-1}$. In parenthesis we give the corresponding $3\sigma$ errors.\\

\end{minipage}
\end{table*}

\subsubsection{NGC~3227}
This nearby Seyfert 1 galaxy is classified as a SAB galaxy and is interacting with NGC~3226. It is known for having a compact (0.4$\arcsec$) double radio source at PA $\sim170\deg$, and
[O\,{\sc iii}] emission extending $\sim7\arcsec$ north-east of the nucleus, in PA $\sim30\deg$ (Mundell et al. 1995), meaning that this object may represent another case in which the radio structure and the [O\,{\sc iii}] are physically misaligned.

Previous reports on high excitation gas in this object are scarce. As in Mrk~3, in spite of the strong cone seen in [O\,{\sc iii}] images, [Fe\,{\sc vii}] in the optical 
and [Si\,{\sc vi}] in the near-infrared (Reunanen et al. 2003; 
Rodr\'{\i}guez-Ardila et al. 2006) are the lines with the highest ionisation
potential reported. 

The STIS data were taken along PA=$-150\deg$, that is, along the
cone axis, where the [O\,{\sc iii}] emission is preferentially extended, and off the
radio jet. The optical spectrum displays weak nuclear emission of [Ne\,{\sc v}], [Fe\,{\sc vii}], and \FeX, not detected in any other aperture. At the distance of
NGC~3727, this implies that the CLR is restricted to the central $\sim$15~pc, making this source
the AGN of the sample with the smallest CLR. This result, moreover,
puts a tighter constraint on the size of the [Fe\,{\sc vii}] gas and agrees with
the values reported by Rodr\'iguez-Ardila et al. (2006). Consistently with the 
optical results here, in the NIR region these authors report that [Si\,{\sc vi}] is the only coronal line detected. It is emitted in a very compact zone, observed only up to 
45~pc N of the nucleus. To the south, no emission is found. It should be
noted that the spectra from Rodr\'{\i}guez-Ardila et al. (2006), both optical
and NIR, were taken with the slit aligned along the radio axis. 
Therefore, it can be concluded that no significant enhancement of the
coronal-line gas is observed in this object due to the local interaction between
gas and the radio jet.

\subsubsection{Mrk~348}

Mrk~348 (NGC~262) is classified as a type 2 Seyfert galaxy. Neff \& de Bruyn (1983) report
that its nuclear radio source consists of a compact core plus two knots
aligned along PA=168$\deg$, with a total size of about 0.15$\arcsec$.
They also report variability at 6 and 21~cm on timescales of months. A
high-resolution narrow-band HST image published by Capetti et al. (1996) shows a
linear structure of narrow-band [O\,{\sc iii}]~$\lambda$5007 emission extended
by 0.45$\arcsec$ at a position angle of $\sim155\deg$. 

Detection of optical coronal lines in this objects has been reported by 
Malkan (1986) and Cruz-Gonz\'alez et al. (1994). They found [Ne\,{\sc v}]
and [Fe\,{\sc vii}], respectively, in the nucleus of this source.
Riffel et al. (2006), using NIR spectroscopy, detected conspicuous emission
of [S\,{\sc viii}], [S\,{\sc ix}], [Si\,{\sc vi}], [Si\,{\sc x}], and [Ca\,{\sc viii}]. 

The HST/STIS data analysed here were taken with the slit at PA=146$\deg$, that is, nearly aligned along the [O\,{\sc iii}] emission. The only coronal lines observed in these spectra are from Ne$^{+4}$. Lines with ionisation potential higher than 100~eV are not observed. As in the case of the lower ionisation lines (i.e. [O\,{\sc ii}], [Ne\,{\sc iii}], and [O\,{\sc iii}]), the \NeV\ emission is very asymmetric. This line is detected up to $\sim$100~pc towards the NW but it is not observed towards the SE beyond the unresolved nucleus. 
As there is no red spectrum available for this object, nothing can be said about the 
[Fe\,{\sc x}] line. 

\begin{table*}
\begin{minipage}{126mm}
\caption{Observed extent of the coronal lines, in parsecs. [O\,{\sc iii}]\l5007 is also included for comparison.}
\label{extension}

\begin{tabular}{lcccccccc}
\hline
& [Ne\,{\sc v}] & [Fe\,{\sc vii}] & [Fe\,{\sc xiv}] & [Fe\,{\sc vii}] & [Fe\,{\sc x}] & [S\,{\sc xii}] & [Fe\,{\sc xi}] & [O\,{\sc iii}]\\
      & \l3425  & \l3760    & \l5303        & \l6086  & \l6374     & \l7611   & \l7892 & \l5007 \\
\hline
Mrk~3    & 230~SW  & --    & --    & 200~SW  &   --  & --    & -- & 470~SW\\
         & 150~NE  & --    & --    & 95~NE   &   --  & --    & -- & 710~NE\\

Mrk~348  & $<30$~SE & -- & -- &  &  &  &  & 250 SE \\
         & 105 NW   &--  &--  & & & & & 610 NW \\

Mrk~573  & 150~NW  & $< 35$~NW  & 85~NW     &      &   120 NW   & &  & 1400~NW\\
      & 120~SE  & 50~SE      & 50~SE    &      &   85 SE   & &  & 1170~SE\\

NGC~1068 & 110~SW & 10~SW  &    & 25~SW   & 10 SW   & 10~SW & 10~SW & 290~SW\\
          & 140~NE & 130~NE &    & 130~NE  & 115 NE  & 20~NE & 50~NE & 375~NE\\

NGC~3081 & 40~SW      & $< 15$~SW & $< 15$~SW   &        &  $< 15$~SW  &         &  & 210~SW\\
      & 55~NE      & 25~NE    & 25~NE        &          & 25~NE  &         &  & 250~NE\\

NGC~3227  & $< 10$~SW &  &  & $< 10$~SW & $< 10$~SW & -- & -- & 50~SW \\
           & $< 10$~NE &  &  & $< 10$~NE & $< 10$~NE & -- & -- & 80~NE \\

NGC~4151 (221$^{\circ}$) & 75~SW   & & & 75~SW & 10~SW  & -- &  & 450~SW \\
	 		 & 70~NE   & & & 30~NE & 10~NE  & -- &  & 255~NE \\

NGC~4151 (70$^{\circ}$)	    & 100~SW   & & & 95~SW & 10~SW  & -- &  & 210~SW \\
   			    & 75~NE    & & & 70~NE & $< 10$~NE  & -- &  & 440~NE \\

NGC~4507 & 130~NW    & 130~NW $^*$  & 35~NW &      &  35 NW  &         &  & 315~NW\\
      & 60~SE     & 60~SE  $^*$  & 35~SE &      &  35 SE  &         &  & 200~SE\\

NGC~5643  & $< 10$ SW & -- & -- &  & -- &  & & 370 SW \\
       & 30 NE & -- & -- &  & -- &  & & 410 NE \\

NGC~7682  & 150 NW & -- &-- & &-- & & & 560 NW\\
       & 50 SE &-- & --& &-- & & & 765 SE\\

\hline
\end{tabular}
\\
\medskip
$^*$ This was measured from the line [Fe\,{\sc vii}]\l3586.\\

\end{minipage}
\end{table*}

\subsubsection{NGC~5643}

This almost face-on barred spiral (Morris et al. 1985)
shows a clear dust lane running along the southern leading edge of the bar to the east. Schmitt, Storchi-Bergmann \& Baldwin (1994) presented [O\,{\sc iii}] and 
H$\alpha$ images, as well as optical spectra of NGC~5643. The images show the
[O\,{\sc iii}] gas to be extended along the east-west direction by ~20\arcsec. East of the nucleus the radio jet (PA $\sim$87$\deg$) lies along the 
southern edge of the ionisation cone (Simpson et al. 1997). 

The HST/STIS optical spectrum of NGC~5643 shows [Ne\,{\sc v}] and [Fe\,{\sc x}]
in the nucleus. The latter feature is, however, very weak. [Fe\,{\sc vii}] is not detected 
at the 3$\sigma$ level in the blue region, while the red spectrum
does not include [Fe\,{\sc vii}]\l6086. However, Morris \& Ward (1988)
report its detection in the optical spectrum of this source.
Overall, the coronal line emission is quite compact and asymmetric. [Ne\,{\sc v}], for instance,
is found up to 30~pc NE while to the SW it is restricted to the unresolved nucleus.
[O\,{\sc iii}]\l5007, in contrast, is significantly more extended to both
NE and SW, observed out to nearly 400~pc.

\subsubsection{NGC~7682}

NGC~7682 is a Seyfert 2 galaxy (Huchra \& Burg 1992) in
interaction with NGC~7683 (Arp 1966). Ionised gas in
H$\alpha$+[N\,{\sc ii}] and [O\,{\sc iii}] is detected on scales of kiloparsecs
in this galaxy (Brodie et al. 1987; Durret 1994). In
the NIR, only K-band spectroscopy was previously reported
by Imanishi \& Alonso-Herrero (2004).
The only optical coronal line observed in the STIS spectrum of NGC~7682 is [Ne\,{\sc v}].
No iron lines or previous reports in the literature of higher ionisation lines in this
wavelength interval were found. In the NIR, however, NGC~7682 shows [Si\,{\sc vi}]~1.963~$\mu$m
and [S\,{\sc viii}]~0.991~$\mu$m (Riffel et al. 2007), whose ionisation potential is 167 and 281~eV, respectively. Though weak, the presence of these latter features indicates that, in a photoionisation scenario, high-energy photons reach the NLR. 

In addition to the unresolved nuclear [Ne\,{\sc v}] emission, this coronal line displays a highly asymmetric extended morphology, observed up to 150~pc NE and 50~pc SW. Due to the low S/N, even for the nuclear spectrum, little can be said regarding the emission line profile. However, nuclear low-ionisation lines are strongly asymmetric, displaying a prominent extended red wing. Beyond the nucleus, the red wing is not observed.

\section{Kinematics of the coronal-line gas}\label{kinematics}

The superb spatial resolution of the STIS spectra allows us to study the 
kinematics of the coronal-line gas at scales down to a few parsecs, and
to compare it with that of lower ionisation gas. We are interested, for instance, in studying if the coronal-line gas is kinematically perturbed and enhanced in regions where the jet 
interacts with the NLR gas, if it is compatible with disk rotation or if it is associated 
instead with outflows. To this purpose we constructed radial velocity maps for those objects that display extended CLRs, with more than one line detected at sufficient S/N. In Figures~\ref{mrk573_vel}--\ref{ngc3081_vel} we show the variation of line centroid velocities and FWHMs of the lines as a function of the distance to the nucleus. The velocity curves measured for [O\,{\sc ii}] and \NeIII\ lines are very similar to the ones of \OIII, hence we did not include them in the plots.
The velocities are given with respect to the systemic velocities of the galaxies derived from their redshifts. Notice that these redshifts, taken from the NASA/IPAC Extragalactic Database (NED), were derived using emission lines, and may have small deviations from the true redshift of the galaxy.
In the cases were a line was fitted by two Gaussian components, the velocities of the more blueshifted line is represented by a filled symbol and the other component by an open symbol. If only one component was detected in a line, the velocity is represented by an open symbol.
When combining high and low spectral resolution data, the FWHM of the lines derived from the former were not used due to the large difference in the spectral  resolutions of both gratings. 

In the following subsections we discuss the results obtained for a subset of galaxies of our sample.

\subsection{Mrk~573}\label{mrk573_kinematics}

\begin{figure}
\includegraphics[width=\columnwidth]{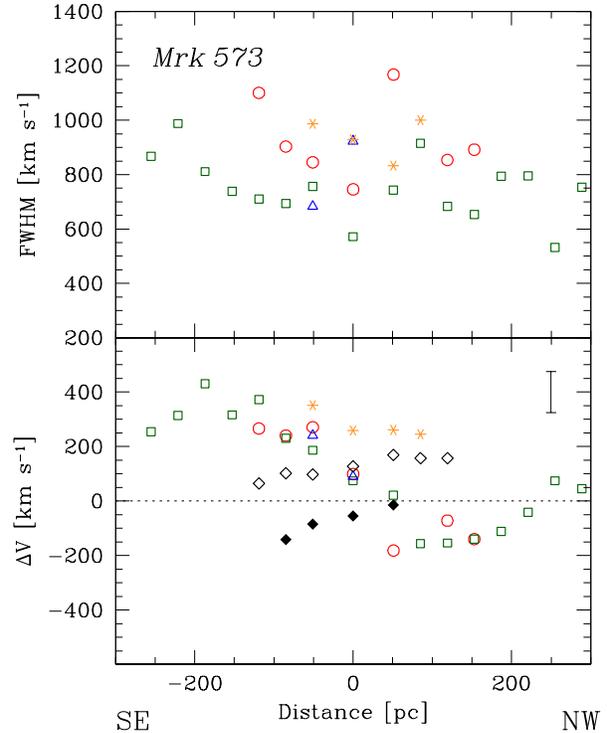}
\caption{Radial velocity (lower panel) and FWHM (upper panel) of the principal lines measured in Mrk~573 as a function of the projected distance to the nucleus. The filled and open symbols represent a component of a line fitted by two gaussian components. Green squares represent \OIII, red circles \NeV, blue triangles \FeVII, black diamonds \FeX, and yellow asterisks \FeXIV. The error bar at the upper right corner of the lower panel shows the maximum error of these measurements.}
\label{mrk573_vel}
\end{figure}

Fig.~\ref{mrk573_vel} shows the variation along the slit of both the FWHM of the most
conspicuous lines studied (upper panel) and the radial velocity for the same set of lines (bottom panel) for this object. Note that the FWHMs of [Fe\,{\sc x}] were not included in the plot because of the large difference in
instrumental resolution between the blue and red spectra. Data for that line, however, 
was employed to study the radial velocity along the spatial direction of the high-ionisation gas. It can be seen that the radial velocity for [O\,{\sc iii}] is in perfect agreement with the results presented by Schlesinger et al. (2009, cf. their fig.~4) using the same data for [O\,{\sc iii}]. The region plotted in Fig.~\ref{mrk573_vel}
maps the high velocity component of the biconical outflow suggested by these authors. 
Our results show that [Ne\,{\sc v}] and probably [Fe\,{\sc vii}], the two coronal lines with the lowest ionisation potential, follow a similar behaviour as [O\,{\sc iii}],
except that the former two are associated with the inner region of the outflow.
[Ne\,{\sc v}] is seen up to only 150~pc to the NW and 120~pc to the SE. The data for
[Fe\,{\sc vii}] are less conclusive as only weak \FeVII\ lines in the blue spectra were available.
Kinematically, [Ne\,{\sc v}] displays a similar rotation curve than 
[O\,{\sc iii}] to the SE. In contrast, at $\sim 50$~pc NW of the 
centre, both lines appear to diverge: the coronal-line gas is shifted to the blue relative to 
[O\,{\sc iii}]. This is consistent with the broadening observed in [Ne\,{\sc v}] at this position. At 85~pc NW a spike is observed at the position of the \NeV\ line, and 
while this line is clearly present, no accurate determination of its shift and FWHM was possible. 

This scenario contrasts to what is seen in [Fe\,{\sc x}]:
It displays a prominent double-peak structure in the region where it is detected {\footnote{Kraemer et al. (2009) recently analysed HST spectra
of the central emission knot of Mrk~573, extracting a single binned 
spectrum from the central 1.1\arcsec. They report a higher FWHM of \FeX\
than seen in any other lines, likely due to the line splitting we
observe in the spatially resolved data (see also our Fig.~\ref{mrk573_fe10})}}.
Moreover, the radial velocity implies that it is not part of the biconical outflow, even though spatially, the region emitting [Fe\,{\sc x}] coincides with that of [Ne\,{\sc v}]. A similar conclusion regarding the rotation pattern can be drawn for [Fe\,{\sc xiv}], which shows a completely different behaviour than the lower ionisation lines. At every distance from the nucleus, the \FeXIV\ line is redshifted with respect to the other lines and the systemic velocity of the galaxy. One possible explanation for this unusual behaviour is that [Fe\,{\sc xiv}] shows similar, but more extreme, line splitting as
\FeX. We however, only detect one peak. Another possibility is some contamination in the red wing of \FeXIV\ by [Ca\,{\sc v}]~\l5309.

The FWHM of the coronal lines varies significantly from point to point. 
For [Ne\,{\sc v}], at the unresolved nucleus, we measured a FWHM of
750~km\,s$^{-1}$ while at 50~pc to the NW it has doubled this value, with
a FWHM of 1200~km~s$^{-1}$, broader than any other NLR feature.

\subsection{NGC~1068}\label{ngc1068_kinematics}

\begin{figure}
\includegraphics[width=\columnwidth]{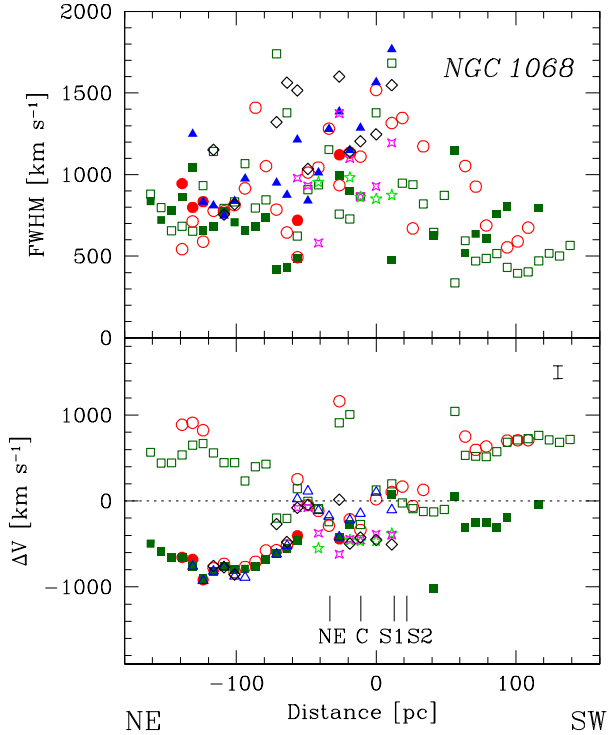}
\caption{Same as Fig.~\ref{mrk573_vel}, but for NGC~1068. Here the magenta four-pointed stars represents \FeXI\ and the green five-pointed stars [S\,{\sc xii}]. The labels in the lower panel indicate the position of the radio knots intersected by the slit.}
\label{ngc1068_vel}
\end{figure}

Many models have been proposed to explain the kinematics of the inner regions of NGC~1068, mainly based on observations of the \OIII\l5007 emission line with high-spatial resolution. Axon et al. (1998) suggested that the kinematics of the NLR is governed by the interaction between the NLR gas and the radio outflow observed in this galaxy. 
Crenshaw \& Kraemer (2000), based on the same set of STIS observations presented here, concluded that a simple kinematic model of a biconical hollowed outflow gives a reasonable explanation for the velocity field of the \OIII\ gas. This model assumes a constant acceleration of the clouds in the inner 140~pc and constant deceleration farther out. 
More recently, Das et al. 2006 re-analysed the radial outflow models and the possible dynamical effects of the radio jet. The models were capable of explaining the large-scale pattern of the \OIII\ emission. However, lateral expansion due to the radio knots (as proposed by Axon et al. 1998) was required to explain the fainter NLR clouds. They concluded that an interaction between the radio jet and the \OIII\ emitting gas is possible, but that these interactions are modest. 

The STIS slit used to observe NGC~1068 was positioned along the small scale radio jet displayed by this galaxy. In Fig.~\ref{ngc1068_vel} we show the rotation curve (lower panel) and FWHM (upper panel) of the coronal lines and \OIII\ measured in NGC~1068 spectra as a function of the distance to the nucleus. The labels indicate the positions where the slit intersects the radio jet knots (see Fig. 2 of Gallimore et al. 1996).

In Fig.~\ref{ngc1068_vel} we can see the extreme complexity of the velocity field of this galaxy. While the \OIII\ line is double peaked in the inner 140~pc, with one component redshifted and another blueshifted, the \FeVII\ and \NeV\ lines are mostly single peaked, following the redshifted \OIII\ component towards the SW of the nucleus and the blueshifted \OIII\ component towards the NE. Exceptions to this are the more external \NeV\ emission regions (from 125~pc to 140~pc NE) where the lines show a second component redshifted by about 1000~km~s$^{-1}$, almost coincident with the redshifted component of \OIII, and a high velocity cloud ($\Delta\rmn{V}=1200$~km~s$^{-1}$) located near the nucleus at 25~pc NE. 
Despite these differences, the overall behaviour of \OIII, \FeVII, and \NeV\ is rather similar, showing a nice symmetric pattern. 
Moreover, the \FeX\ line also seems to follow the blueshifted component of the \OIII\ line at the NE side of the nucleus. At the nucleus and 10~pc SW the \FeX\ is blueshifted with respect to the lower ionisation emission lines, with $\Delta\rmn{V}\sim 500$~km~s$^{-1}$. 
On the other hand, the higher ionisation \FeXI\ and [S\,{\sc xii}] lines are slightly blueshifted with respect to lower ionisation emission lines, and seem to share the same overall kinematics. This blueshift is about 500~km~s$^{-1}$ with respect to the systemic velocity of the galaxy, except for the \FeXI\ line emitted at 40~pc and 50~pc NE which has almost the same velocity as the galaxy, as is the case for the \FeX\ and \FeVII\ lines.  

In the upper panel of Fig.~\ref{ngc1068_vel} we show the variation of FWHM with the distance from the nucleus. These values cover a wide range of velocities, from about 300\kms\ to $\sim 1800$\kms. No clear correlation between the FWHM and IP of the lines is observed.

We can see in Fig.~\ref{ngc1068_vel} a strong similarity between the velocity curves of the CLs and \OIII, suggesting a common origin. Moreover, the velocity curves and FWHMs of the lines are not especially perturbed at the positions of the radio knots. This implies that the radio jet does not have a significant local influence over the coronal and \OIII\ emission-line gas kinematics.

\subsection{Mrk~3}\label{mrk3_kinematics}

\begin{figure}
\includegraphics[width=\columnwidth]{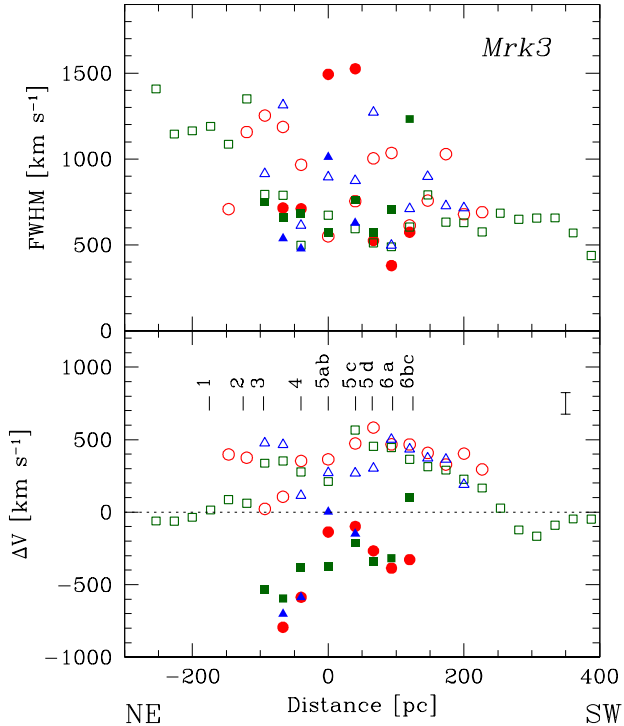}
\caption{Same as Fig.~\ref{mrk573_vel}, but for Mrk~3. The labels indicate the position of the radio knots intersected by the slit. }
\label{mrk3_vel}
\end{figure}

The STIS slit used in the observations of Mrk~3 coincides with the position of the radio jet reported by Kukula et al. (1993). Therefore, it is another excellent opportunity to study the small scale relation between the radio and the coronal line emission and its influence on the kinematics of the gas. In Fig.~\ref{mrk3_vel} we present the radial velocities (lower panel) and FWHMs (upper panel) of the principal lines detected in this galaxy as a function of the distance to the nucleus. The labels in the lower panel indicate the position of the radio features intersected by the slit (see Fig.~2 of Kukula et al. 1999).

In the lower panel of Fig.~\ref{mrk3_vel} we can see that the inner regions of Mrk~3 display line profiles splitted into two distinct velocity systems, one redshifted and the other blueshifted with respect to the systemic velocity of the galaxy. Outwards (more than 120~pc SW and 100~pc NE) only the redshifted component is observed. 
There is a general trend of the line-emitting gas to accelerate as one moves away from the nucleus, it reaches a maximum velocity at about 100~pc from the centre, and then decelerates back to the systemic velocity of the galaxy 100~pc farther out. The maximum redshift of the [Ne\,{\sc v}] and [Fe\,{\sc vii}] lines, $\Delta\rmn{V}\simeq 550$~km~s$^{-1}$, is achieved at 70~pc SW and 90~pc SW, respectively. The maximum blueshift of these lines is reached at 65~pc NE from the nucleus and is slightly higher than the redshift ($\Delta\rmn{V}\simeq 750$~km~s$^{-1}$). 
As in the case of NGC~1068, the velocity structure and FWHM of the lines do not seem to be correlated with the positions of the radio jet knots. However, the line splitting is observed almost all along the fraction of the jet probed by the slit. Moreover, no trend is observed in the FWHMs of the lines, which strongly vary with the distance from the nucleus, from $\sim 400$~km~s$^{-1}$ up to the extreme value of the \NeV\l3425 line of $\rmn{FWHM} \sim 1500$~km~s$^{-1}$ in the nucleus and 40~pc to the SW.

Previous kinematic models for the [O\,{\sc iii}]\l5007 emitting gas of Mrk~3 were analysed by Ruiz et al. (2001) based on the same set of data used in this work together with [O\,{\sc iii}] slitless spectra. The best fit model indicates that the NLR gas is located in a partially filled bicone and is accelerated radially away from the nucleus and then decelerates at a constant rate. Another kinematic model was proposed by Cappeti et al. (1999), in which the gas forms part of an expanding coocon of gas circumscribing the radio jet.

The similar behaviour presented by the CLs and [O\,{\sc iii}] suggests that the [Ne\,{\sc v}] and [Fe\,{\sc vii}] gas is also part of the same outflow that governs the lower ionisation gas in the central 400~pc of Mrk~3. However, some deviations from the \OIII\ emission-line gas are present, mainly towards the NE. At a distance of 70~pc and 90~pc NE from the nucleus, the \NeV\ lines are blueshifted with respect to the \OIII\ lines by an amount of $\Delta\rmn{V}\sim 400$~km~s$^{-1}$, whereas further out the trend reverses and the \NeV\ lines are redshifted with respect to the lower ionisation lines.

\subsection{NGC~4151}\label{ngc4151_kinematics}

High-resolution spectroscopy based on HST data has been
employed by several authors to study NGC~4151. 
The kinematics derived from HST long slit observations of \OIII\ 
(Nelson et al. 2000; Crenshaw et al. 2000) show evidence of
three components: a low velocity system, consistent with
normal disk rotation, a high velocity system in radial outflow 
at a few hundred \kms, and an additional high velocity system 
with velocities up to 1400\kms, as previously found from 
STIS slitless spectroscopy (Hutchings
et al. 1998, 1999; Kaiser et al. 1999). The general consensus
points to signatures of a radial outflow in the form of a wind, 
with no interaction with the radio jet. 

\begin{figure}
\includegraphics[width=\columnwidth]{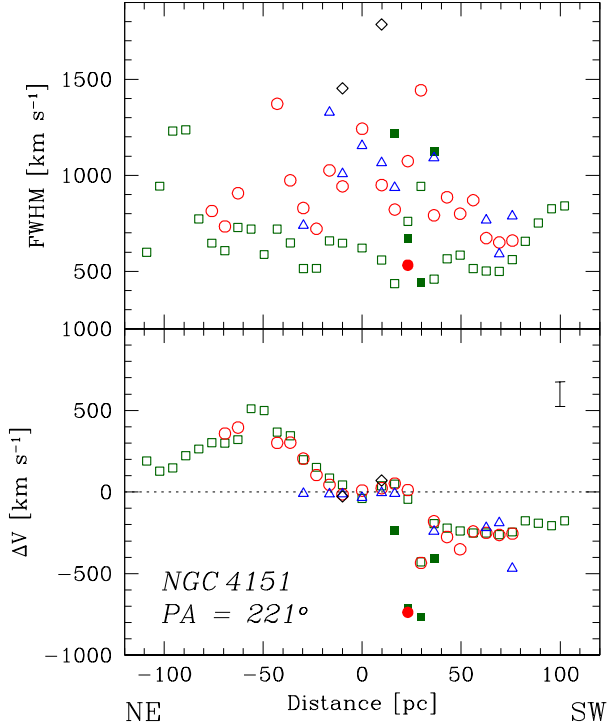}
\caption{Same as Fig.~\ref{mrk573_vel}, but for NGC~4151 at PA = 221$^{\circ}$.}
\label{ngc4151_vel}
\end{figure}

\begin{figure}
\includegraphics[width=\columnwidth]{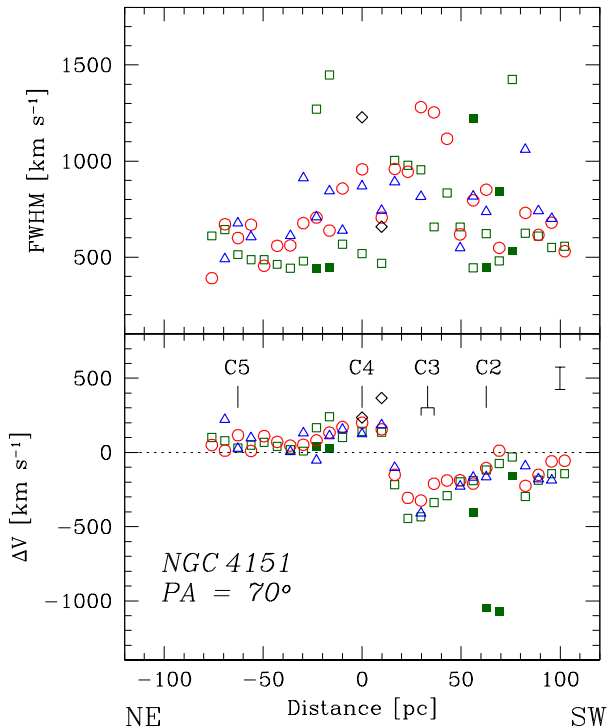}
\caption{Same as Fig.~\ref{mrk573_vel}, but for NGC~4151 at PA = 70$^{\circ}$. The labels in the lower panel indicate the position of the radio knots intersected by the slit.}
\label{ngc4151_vel_70}
\end{figure}

Overall, the rotation curves found for [O\,{\sc iii}], shown in the 
lower panel of Fig.~\ref{ngc4151_vel} and Fig.~\ref{ngc4151_vel_70}, are consistent with 
that derived by Crenshaw et al. 2000. Along PA=$221\deg$, a velocity
gradient of over 500~km~s$^{-1}$ in the inner 50~pc NE of the nucleus
is detected, with all lines being redshifted. The gas first accelerates to a
maximum velocity of 500\kms\ relative to the systemic velocity of the
galaxy. It then decelerates back to about 100\kms\ at $\sim$100~pc.
Towards the SW, at $\sim$30~pc from the centre, a double-peak structure is 
detected, with the red peak accompanying the systemic velocity of
the galaxy and the blue peak shifted by $\sim$700\kms\ to the blue from the former. 
The relative separation between both
peaks decreases in the next few parsecs outwards. At $\sim$45~pc from the
centre, only the red peak is detected. From that point, it suffers a slight acceleration 
to the blue, reaching 250\kms\ at 70~pc, and then decelerates to 200\kms\ at 100~pc. 
Along PA=$70\deg$ the velocity curve is more flat, deviating very little from the systemic velocity of the galaxy. At $\sim 25$~pc SW the lines reach a maximum blueshift of $\sim 400$\kms. Further out, at about 65~pc SW, a second peak in the \OIII\ lines is measured, with a blueshift of $\sim 1000$\kms, not seen in the CLs. This splitting of the lines coincides with the component C2 of the radio jet.
The highest ionisation line detected in NGC~4151 is [Fe\,{\sc x}]\l6374. When detected, this line has the same velocity as the lower ionisation gas mapped by \OIII, \NeV, and \FeVII.
In the lower panel of Fig.~\ref{ngc4151_vel_70} we marked the positions where the slit intersects the radio knots. From this, it is clear that there is no correlation between the radio emission and the CL kinematics, suggesting that the radio jet has no significant influence on the overall kinematics of the gas emitting the CLs.

The variation of FWHM with the distance from the nucleus is 
plotted in the upper panel of Fig.~\ref{ngc4151_vel} and Fig.~\ref{ngc4151_vel_70}. It is easy to see that there is a large scatter in FWHM from point to point, both for lines of the same element and among lines of different species. 
Overall, the scatter in FWHM among the 
different ions is the largest in the region where double-peaked lines were 
detected. This may imply that a number of lines are actually double-peaked, but no longer spectroscopically resolved.

Our radial velocity curves derived from the STIS data which includes, for the
first time, the coronal lines [Ne\,{\sc v}], [Fe\,{\sc vii}], and \FeX\  
(see Fig.~\ref{ngc4151_vel} and Fig.~\ref{ngc4151_vel_70}) show that these CLs follow the kinematics traced by [O\,{\sc iii}]. This implies that they follow the same velocity field and originate from the outflow itself, with no important influence of the radio jet.

\subsection{NGC~4507}

\begin{figure}
\includegraphics[width=\columnwidth]{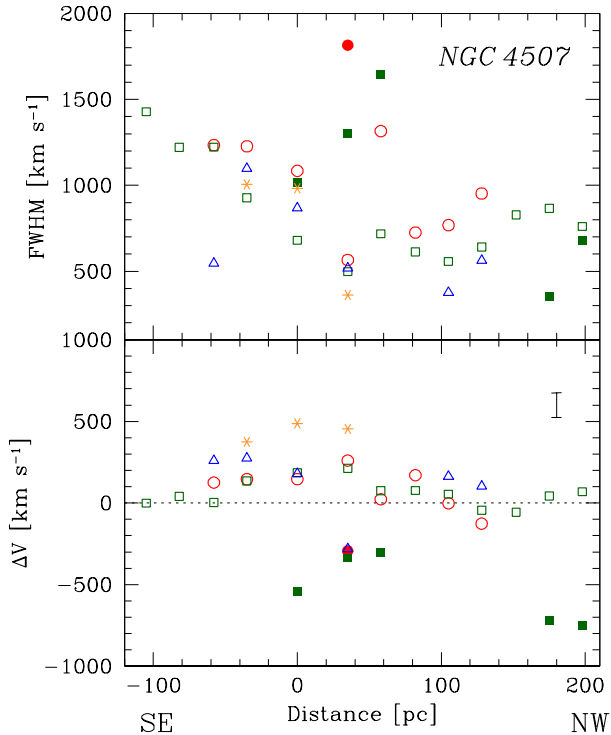}
\caption{Same as Fig.~\ref{mrk573_vel}, but for NGC~4507.}
\label{ngc4507_vel}
\end{figure}

Fig.~\ref{ngc4507_vel} shows the radial velocity field (lower panel) and FWHM (upper panel) of the principal lines measured for NGC~4507 along the slit. 
The velocity curve of this galaxy is very different to the ones studied in the previous sections, showing a rather flat pattern with velocity shifts closer to the systemic velocity of the galaxy. 
At the nucleus, the \OIII\ line displays a double-peaked profile, with one component redshifted by $\Delta\rmn{V}=200$~km~s$^{-1}$ and another blueshifted by $\Delta\rmn{V}=550$~km~s$^{-1}$. 
At this point the coronal lines of \NeV\ and \FeVII\ show a small blue asymmetry, but the two components are not resolved. The centre of these lines coincides with the red component of \OIII. The maximum of this asymmetry is reached at $\sim 35$~pc NW, where the strongest lines (i.e. \OIII\ and \NeV) can be characterised by the sum of two gaussians separated by $\Delta\rmn{V}\sim 600$~km~s$^{-1}$. One component describes a strong narrow core ($\rmn{FWHM}\sim 500$~km~s$^{-1}$) and the  second component represents a blueshifted wider ($\rmn{FWHM}\sim 1500$~km~s$^{-1}$) part. Notice that the \FeVII\ velocity at this point coincides with the ones of the blueshifted components of \OIII\ and \NeV. The blue asymmetry smoothly vanishes as one moves away from the nucleus and a tenuous red wing appears in the emission lines emitted in the outer regions. Further out, where no coronal lines are observed, the \OIII\ lines in the NW side of the nucleus are double peaked, showing a strong blueshift that reaches 750~km~s$^{-1}$.
In general, we can see that the \NeV\ and \FeVII\ emitting gas share the same kinematics with the \OIII\ gas, with a weak tendency of the \FeVII\ lines to be redshifted with respect to the other two lines. 
On the other hand, the \FeXIV\ line is redshifted with respect to the lower ionisation lines, with a maximum shift of 500~km~s$^{-1}$ at the nucleus. As we mentioned in Section~\ref{mrk573_kinematics}, the possible presence of weak [Ca\,{\sc v}]~\l5309 on the \FeXIV\ red wing can be, in part, the reason for the measured shift.

The higher resolution red spectra show lines with much more complex structures. As this spectrum was taken with the slit positioned in a different angle than the blue spectrum, we do not include the \FeX\ line in Fig.~\ref{ngc4507_vel}. As we mentioned in Section~\ref{results_4507}, two components were required to describe the nuclear \FeX\ line: one low velocity, broad component ($\Delta\rmn{V}=155$~km~s$^{-1}$, $\rmn{FWHM}=685$~km~s$^{-1}$) and a narrow component ($\rmn{FWHM}=130$~km~s$^{-1}$) redshifted by 350~km~s$^{-1}$ with respect to the systemic velocity of the galaxy. 
These components are consistent with the velocity-shifts and FWHMs measured for the higher velocity [O\,{\sc i}] lines, the latter being slightly broader than the former. No counterpart for the more intense [O\,{\sc i}] component was detected in \FeX. 
At both sides of the nucleus, the \FeX\ line profiles were described by one gaussian component  with velocities redshifted by 250~km~s$^{-1}$ and $\rmn{FWHM} = 550$~km~s$^{-1}$. 

\subsection{NGC~3081}

\begin{figure}
\includegraphics[width=\columnwidth]{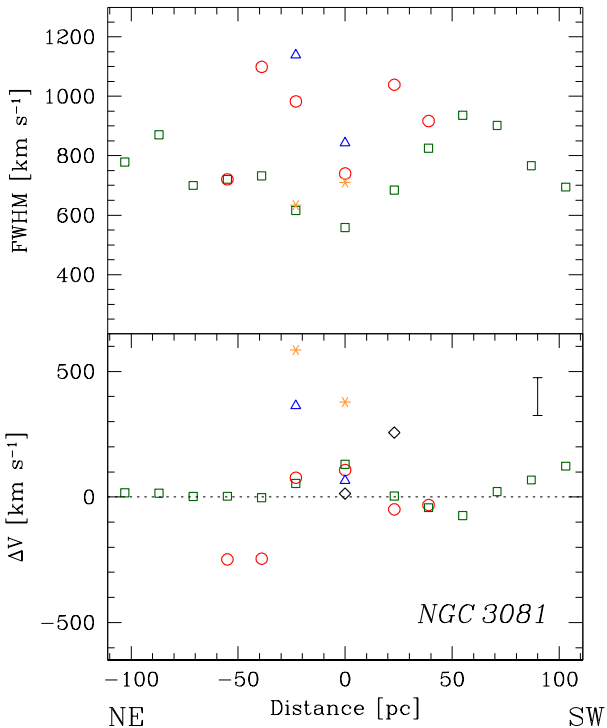}
\caption{Same as Fig.~\ref{mrk573_vel}, but for NGC~3081.}
\label{ngc3081_vel}
\end{figure}

Very little information exists in the literature on the kinematics of the spatially resolved NLR gas of NGC~3081 (Ruiz et al. 2005). Here, we used the STIS spectra to map the kinematics of the \OIII, \NeV, \FeVII, \FeX, and \FeXIV\ emission-line gas in the inner 200~pc of this galaxy. 

As in the case of NGC~4507, the rotation curve of NGC~3081 (lower panel of Fig.~\ref{ngc3081_vel}) is fairly flat. The velocity of the \OIII\ emitting gas is no greater than a few kilometres per second, with a maximum of $\sim 150$~km~s$^{-1}$ at the nucleus. The FWHMs of the \OIII\ line are very similar, with little variation along the spatial direction. 
The velocity dispersion of the coronal emission lines is slightly higher, with the lines either blueshifted or redshifted with respect to the systemic velocity of the galaxy. At the nucleus, all lines but \FeXIV\ have the same shift of the line peak, corresponding to the systemic velocity of the  galaxy. At this point, the \FeXIV\ line shows a receding velocity of $\sim 400$~km~s$^{-1}$. Next to the nucleus ($\sim 25$~pc NE and SW), the gas producing lines with $\rmn{IP}>100$~eV is redshifted by more than $\Delta\rmn{V}=250$~km~s$^{-1}$. The maximum velocity is achieved by the \FeXIV\ emitting gas, which presents a  $\Delta\rmn{V}=600$~km~s$^{-1}$. Further out, at $\sim 45$ and 55~pc NE, the \NeV\ lines are blueshifted by $\Delta\rmn{V}=250$~km~s$^{-1}$. 

If we take into account that NGC~3081 is a galaxy observed nearly face-on, gas in a rotating disk aligned with the equatorial plane of the galaxy will be observed with the same systemic velocity of the galaxy. This seems to be the case of the \OIII\ emission-line gas.  
On the other hand, the higher dispersion of velocity and FWHM shown by the coronal-line gas suggest that it is governed by different kinematics than the \OIII\ emitting gas. The CLR of this AGN is rather turbulent and chaotic.
Moreover, no coincidence in the FWHM or peak velocity between coronal lines of similar ionisation potential is found, indicating that they are produced in different regions along the line of sight.

\section{Demystifying the coronal-line region of active galactic nuclei}

The high-spatial resolution achieved with STIS allowed us to study not
only the global characteristics of the optical coronal line emitting gas but also, for the first time, its small scale properties as a function of the distance to the active nucleus. In this section we discuss the general results that are obtained from the analysis of these data, including the location of the coronal line emitting region, the origin of this emission, and its relation with the radio jets.

\subsection{Summary of key results}

\begin{figure}
\includegraphics[width=\columnwidth]{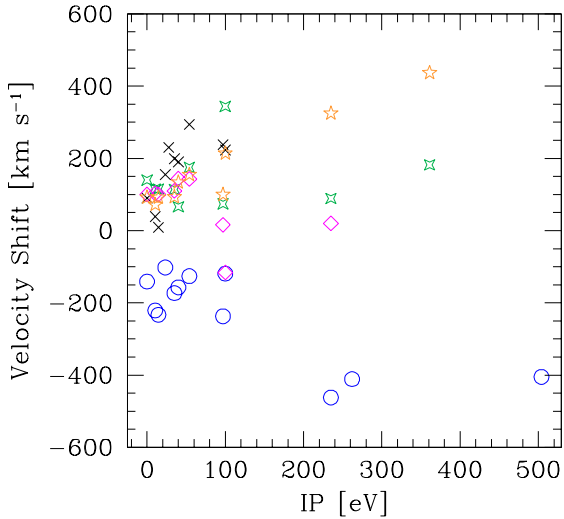}
\caption{Velocity shift of the lines measured in the integrated spectra of the galaxies as a function of the IP. The black crosses correspond to the lines measured in Mrk~3, green four-point stars to Mrk~573, blue circles to NGC~1068, magenta diamonds to NGC~3081, and yellow five-point stars to NGC~4507.}
\label{IP_corr}
\end{figure}

The spectra analysed in this work revealed multi-component and very complex CLRs, showing a clumpy morphology similar to the one displayed by the low-ionisation lines. The most extended CLs, \NeV\ and \FeVII, are observed from a few parsecs up to hundreds of parsecs from the nucleus. The {\em most compact} CLR is observed in NGC~3227, where the \NeV\ and \FeVII\ extent is restricted to the central $\sim$15~pc. On the other hand, Mrk~3 exhibits the {\em largest} \NeV\ and \FeVII\ emission regions, with a total extent of 375~pc and 300~pc, respectively. The higher ionisation lines (with $\rm{IP} >100$~eV), \FeX, \FeXI, \FeXIV, and [S\,{\sc xii}], are emitted in more compact regions. \FeX\ is the most extended of this group of lines, varying from 15~pc up to 200~pc, while the other lines are always observed much closer to the nucleus, in regions of less than 100~pc diameter. Broadly speaking, a stratification is observed, in the sense that lines with higher ionisation potentials are more concentrated towards the centre. One interesting aspect is that some objects show a lack of coronal lines with IP larger than 100~eV
even though they display strong and extended soft X-ray emission.

From the analysis of the kinematics of the emission-line gas, we found that the CLs tend to follow the same velocity distribution as \OIII, in particular the \NeV\ and \FeVII\ lines. Line splitting is sometimes observed in coronal and lower ionisation lines (e.g. \NeV, \FeVII, \FeX, and \OIII). In those cases where \FeXIV\ was observed, the line centre was redshifted with respect to the lower ionisation lines. This means that no universal trend between line shift and IP is observed when high-spatial resolution is employed.
This is in contrast with previous reports from low-spatial resolution spectroscopy, usually integrating the {\em whole} NLR, where the peak of the CLs tend to be more blueshifted 
than lower ionisation lines (e.g., De Robertis \& Osterbrock 1984; Appenzeller \& \"{O}streicher 1988; De Robertis \& Shaw 1990; Marconi et al. 1996; Reunanen et al. 2003). In order to check if the same trend is also apparent in our particular galaxies, we integrated the light distribution along the slit mimicking a seeing-limited observation. We then measured the shift of the central peak of the lines and plotted them as a function of IP as shown in Fig.~\ref{IP_corr}. 
It can be seen that in Mrk~573 and NGC~3081 no correlation is present at all, while in NGC~4507 the shift of the lines increases with the IP. The trend shown by the latter was previously reported by Appenzeller \& \"{O}streicher (1988).
Only in NGC~1068 a tendency for high-ionisation lines to be more blueshifted than lower ionisation ones can be confirmed. 

Three of the galaxies in the sample (NGC~4151, NGC~1068, and Mrk~3) were observed with the slit along the nuclear radio jets they exhibit. From the comparison of the radio and line emission we generally found no enhancement of the coronal line emission, or perturbations of its kinematics, at the positions of the radio knots.

\subsection{Origin of the coronal line emission}

Two key types of models might explain the coronal lines and their properties: either photoionisation by the central continuum source (e.g., Korista \& Ferland 1989; Oliva et al. 1994; Ferguson et al. 1997a; Binette et al. 1997; Nagao et al. 2003; Mullaney et al. 2009),
or shock ionisation or other processes directly or indirectly linked to the
presence of radio jets (e.g., Tadhunter et al. 1988; Viegas-Aldrovandi \& Contini 1989; Morse, Raymond \& Wilson 1996; Axon et al. 1998).

We discuss these possibilities in turn.
In the context of photoionisation, important parameters which
would affect the measured coronal line spectrum are the
shape and strength of the ionising continuum [either from
a central ``point source'' and/or including spatially extended
emission as it has been observed in several nearby Seyfert
galaxies (e.g., NGC~1068: Young et al. 2001; NGC~4151: Ogle et al. 2000)],
the gas density, the cloud column densities (matter-bounded
versus ionisation-bounded), and the metal abundances,
which however, should have no strong local variations. 

The presence of radio jets could affect the emission-line ratios and kinematics in
several major ways: 
firstly, by collisional ionisation
from local shocks directly originating from jet-cloud interactions.
In this scenario gas is heated up to temperatures of several $10^6$~K and a close spatial coincidence between jets and gas clouds is expected. 
Secondly, shocks can compress the gas which in turn is photoionised by the central source (Viegas-Aldrovandi \& Contini 1989). In this case a close link between the jet and the morphology and kinematics of the emission-line gas is expected. 
Thirdly, jet-cloud interactions could lead to fast auto-ionising shocks (Dopita \& Sutherland 1995, 1996), photoionising surrounding gas. 
A grid of models taking into account ionisation by the central source, the diffuse radiation of shocked and ionised gas, and collisional ionisation were presented by Contini \& Viegas (2001). The application of these models to AGNs shows that indeed, fast shocks could be responsible for a fraction of the observed line flux.
Finally, jet-gas interaction might produce cocoons and lateral flows, affecting gas at larger distances than directly at the loci of the jet axis (Taylor, Dyson \& Axon 1992; Steffen et al. 1997). 

The physics of jets and their interaction with the surrounding medium is highly complex. Moreover, this scenario is even more complicated when we also consider the additional effect of ionising photons from the central continuum source. Jets might have indirect effects on the local gas physics, even if photoionisation is the dominant ionisation source; e.g. by locally enhancing the gas density. Such a case is difficult to rule out, since the presence of the jet would mostly add to the spectral complexity of the line emission, without directly accounting for all its features.

\subsubsection{Coronal lines and their relationship with radio jets}\label{CL-jet}

The tendency of photoionisation models to underpredict the highest-ionisation CLs, and the spatial coincidence between the radio and high-ionisation emission, led several authors to suggest that fast shocks driven by radio jets can play an important role in the formation of CLs (e.g., Axon et al. 1998).
In order to address this point we selected from the initial sample of objects three galaxies, Mrk~3, NGC~1068, and NGC~4151, which show a strong, highly collimated small-scale radio jet. The HST/STIS data of these galaxies are particularly suitable to test the hypothesis of shocks as precursors of the CLs since the slits used in the observations were nearly aligned with the radio jets. 

In Sections~\ref{ngc1068_results}, \ref{mrk3_results}, and \ref{ngc4151_results}  we described the main properties of NGC~1068, Mrk~3, and NGC~4151, and the results obtained from the measurements of their coronal-line fluxes. We showed that, in spite of the similarities in the radio emission displayed by these galaxies, there is a notable difference between them: while NGC~1068 displays CLs with a wide range of ionisation potentials, from 99~eV up to 505~eV, NGC~4151 only shows coronal lines of IP $< 300$~eV. The case of Mrk~3 is even more critical as the only coronal lines detected are those of IP~$\sim 100$~eV.
Moreover, the intensity distribution of these lines shows a similar morphology than the lower ionisation lines of \OIII\ and \NeIII. Furthermore, there is no enhanced emission at the positions where the radio knots are observed. In particular, NGC~4151 was observed at two position angles, one along the radio jet and the other 30 deg away from the radio axis. The line intensities and line ratios are remarkably similar at these two positions, not showing any signs of increased emission along the position of the radio jet, as noticed by Nelson et al. (2000). 
Additionally, the kinematics of the CLs (Section~\ref{ngc1068_kinematics} for NGC~1068, Section~\ref{mrk3_kinematics} for Mrk~3, and Section~\ref{ngc4151_kinematics} for NGC~4151) 
do not seem to be locally affected by the radio jet. In none of the three cases we see a correlation between the position of the radio knots and the velocity field or the FWHM of the lines. Although several models have been proposed to explain the velocity curves of these galaxies, including models involving jet-cloud interaction, the simple model of a radial outflow (Nelson et al. 2000; Crenshaw \& Kraemer 2000; Ruiz et al. 2001; Crenshaw et al. 2000) was able to successfully reproduce the kinematics of \OIII\ for the three galaxies and, therefore, also the kinematics of the CLs. 

Additional information about the jet-CL interaction is provided by the galaxy Mrk~573. This galaxy also harbours a strong radio jet in its nucleus, but unfortunately the STIS slit was positioned in another direction and does not intersect the radio knots. However, it can still give us useful information. 
Mrk~573 represents an extreme case of \FeX\ emission. The coronal line emission is strongly peaked, with the maximum in the nucleus and no relation at all with the knots and arcs seen in the WFPC2 F606W filter image by Pogge \& Martini (2002). This scenario is reinforced by Whittle et al. (2005) and Schlesinger et al. (2009), who presented detailed calculations of the radio jet energetics and its modest impact on the gas ionisation. They both showed, for Mrk~78 and Mrk~573 respectively, that the central source is the main driver for the observed luminosity, while the mechanical luminosity associated to the outflow accounts only for a few percent of the estimated bolometric luminosity.

\begin{figure}
\includegraphics[width=\columnwidth]{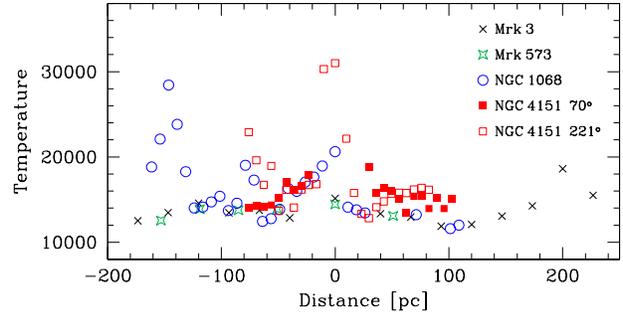}
\caption{Gas temperature measured from the \OIII\ lines for Mrk~3, Mrk~573, NGC~1068, and NGC~4151.}
\label{temperature}
\end{figure}

As we mentioned above, in some cases, shocks can lead to very high temperatures in the line-emitting gas. In order to check the temperatures, we employed the standard diagnostic \OIII~(4958+5006)/4363 line ratio to determinate the temperature of the O$^{+2}$ gas as a function of the distance to the nucleus (Fig.~\ref{temperature}). We found temperatures ranging from $\sim 10000-20000$~K at most for the positions where the ratio could be determined. The temperatures derived for the inner regions of NGC~4151 are probably overestimated due to a strong blend of the \OIII\l4363 line with the broad line of H$\gamma$.
Temperatures lower than 20000~K are typical of photoionised gas, and do not indicate any additional energy input source, as would be the case of shocks strongly interacting with the gas. Although these temperatures were measured for the \OIII\ region, the similarity between the CLs and lower ionisation gas suggests a common origin and therefore no important contribution of shocks to the CLs.

From all the above, we see that there is no clear evidence of a {\em local} influence of the radio jet on the coronal-line gas. However, there is the possibility of the presence of hot cocoons generated by the radio jet that affect the gas located at larger distances from it. In this scenario, the velocity curves are expected to form a broken ellipsoid, with the approaching gas strongly blueshifted and the receding gas strongly redshifted, independent of the observation angle. We can discard the presence of an expanding cocoon in the case of NGC~4151, since the lines displayed by this galaxy have only one component. 
For NGC~1068 and Mrk~3 this scenario can not be tested with the data presented here. However, integrated emission-line fluxes from long slit spectroscopy (Koski 1978) show that, in the case of Mrk~3, the ratio of \FeVII/\OIII\ is very similar to that measured from the STIS data, implying that if cocoons have any impact on the coronal line emission it should be quite small.

Although we can not completely discard the contributions of shocks to the formation of the CLs, the lack of correlation between the positions of the radio knots and the flux distribution and kinematics of the CLs in these galaxies, and the different ionisation ranges displayed by them, suggest that the interaction of the radio jet with the surrounding gas does not have a major effect in the emission of the CLs locally. However, with the data presented here, we can not exclude the presence of expanding cocoons formed by the radio jet in NGC~1068 and Mrk~3.  Further progress can be achieved from the analysis of 3D spectroscopy, mapping the emission-line gas not only along the radio jet but also the gas located a larger distances from it.

\subsubsection{Predictions from photoionisation modelling}\label{photo}

\begin{figure*}
\includegraphics[]{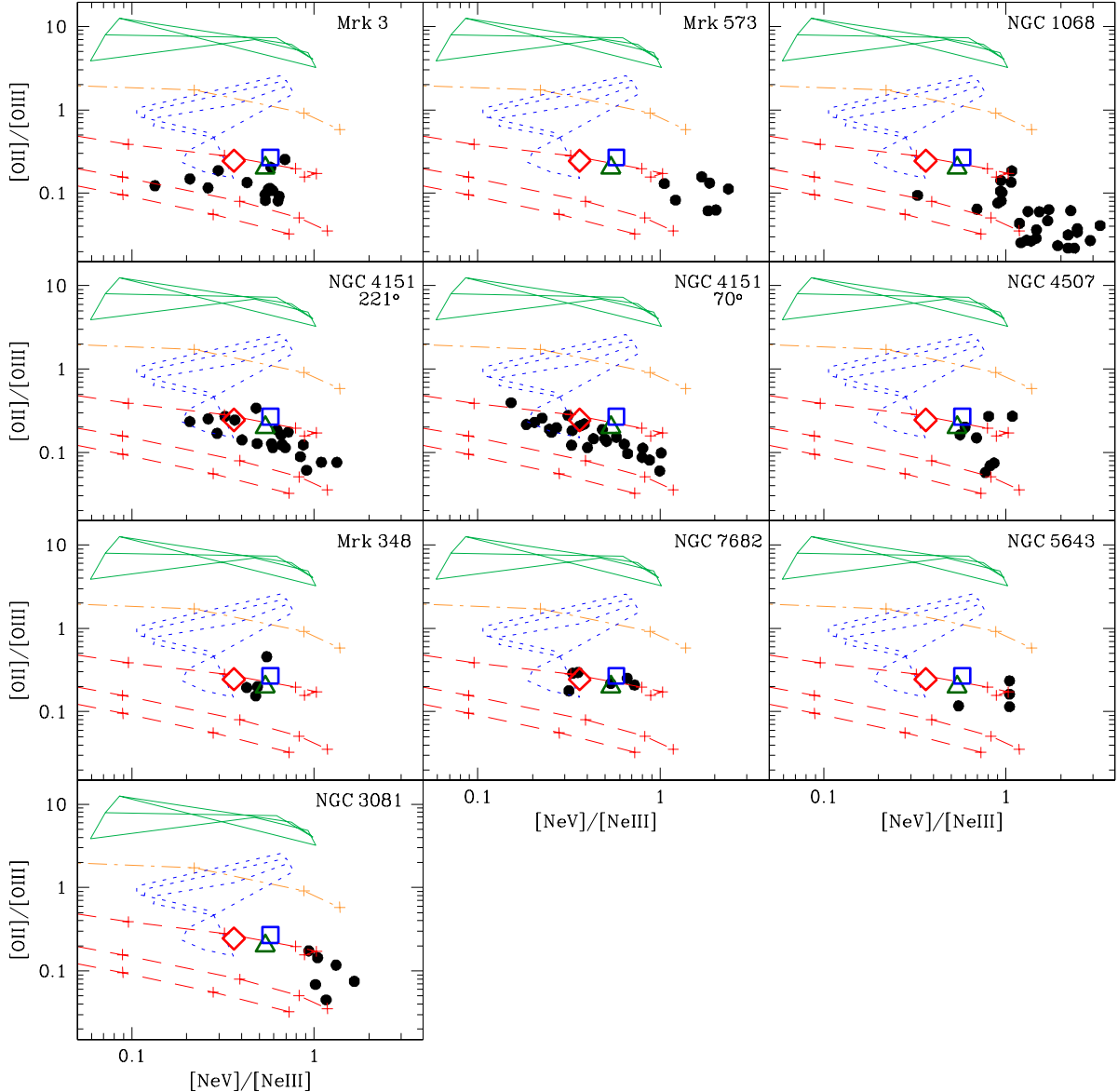}
\caption{[O\,{\sc ii}]/\OIII\ versus \NeV/[Ne\,{\sc iii}] line ratios diagram. The points correspond to the line ratios measured in the galaxies of the sample at different distances from the nucleus. Red pluses linked by dashed-lines correspond to the predictions of optically-thick single-slab power-law photoionisation models with spectral indices of $-1.0$, $-1.5$, and $-2.0$ (from top to bottom), and a sequence in the ionisation parameter covering the range $5\times10^{-4}< U < 10^{-1}$, increasing from left to right (see fig.~4 of Tadhunter 2002). Orange pluses linked by a dot-dashed line correspond to the predictions of photoionisation models including matter-bounded clouds of Binette, Wilson \& Storchi-Bergmann  (1996), with the A$_{\rm{M}/\rm{I}}$ parameter covering the range $10^{-4} < \rm{A}_{\rm{M}/\rm{I}} < 1$, increasing from left to right. Green solid lines and blue dotted lines correspond to pure shocks and ``50\% shock $+$ 50\% precursor'' models from Dopita \& Sutherland (1996), respectively. Each sequence corresponds to a fixed magnetic parameter ($B/\sqrt{n} = 0,1,2,4$~$\mu$G~cm$^{-3/2}$) and a shock velocity $v_s$ varying between 150 and 500\kms. The green triangle corresponds to Ferguson et al. (1997a) models. The blue square and red diamond represent the BB20 and PB50 models of Komossa \& Schulz (1997), respectively.}
\label{models}
\end{figure*}

\begin{figure*}
\includegraphics[]{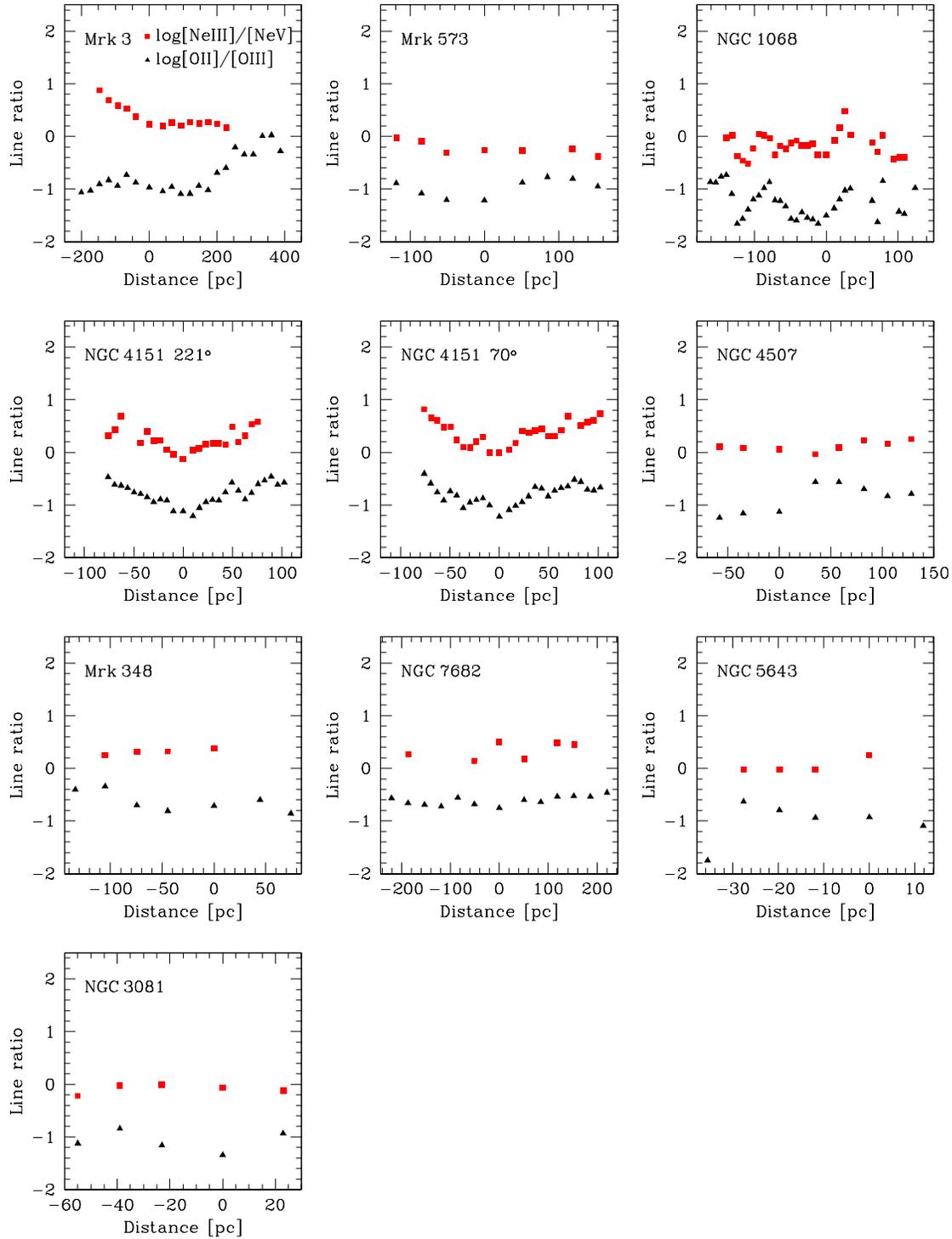}
\caption{[O\,{\sc ii}]/\OIII\ (black triangles) and [Ne\,{\sc iii}]/\NeV\ (red squares) line ratios as a function of the distance to the nucleus of the galaxies in the sample.}
\label{U}
\end{figure*}

\begin{figure*}
\includegraphics[width=0.9\textwidth]{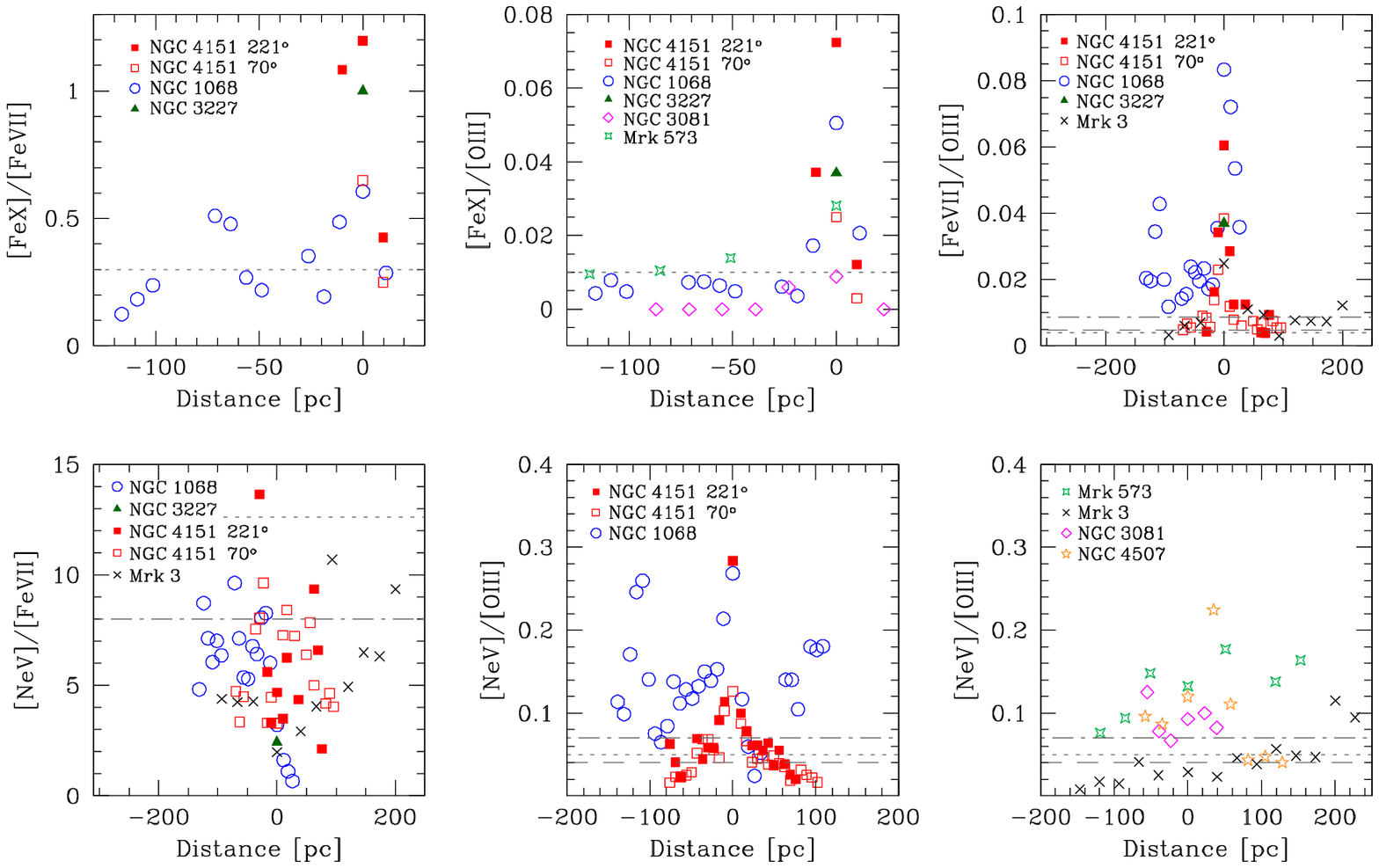}
\caption{Coronal emission-line ratios measured for the galaxies in the
sample as a function of the distance to the nucleus. Dotted lines correspond to the predictions of Ferguson et al. (1997a) models. Dot-dashed and dashed lines correspond to the BB20 and PB50 models of Komossa \& Schulz (1997), respectively. Notice that, for visualisation purposes, we splitted the sample of galaxies when plotting the \NeV/\OIII\ ratio.}
\label{fe7fe10}
\end{figure*}

In this Section, we compare the observed emission-line ratios with photoionisation models. In a first step, we compare model predictions for the low-ionisation line ratio [O\,{\sc ii}]/\OIII\ versus that of [Ne\,{\sc iii}]/\NeV. In a second step, we focus on the higher ionisation CLs, for which fewer model predictions are available.

Two model approaches have been followed in the past. On the one hand,
authors have attempted to match CL strengths in {\em samples} of Seyfert
galaxies, with an eye on the highest observed line ratios of CLs relative
to low-ionisation lines (e.g., Erkens et al. 1997; Komossa \& Schulz 1997;
Binette 1998; Rodr\'iguez-Ardila et al. 2002; Nagao et al. 2003;
Rodr\'igue-Ardila et al. 2006). On the other hand, attention has focused
on individual well-studied Seyfert galaxies which have spatially resolved
NLRs, and photoionisation models have been presented which successfully
reproduce the emission line ratios (with focus on low-ionisation lines,
but including \NeV\ and \FeVII) of individual clouds. For the STIS
spectra analysed here, this was done for NGC~4151 (Nelson et al. 2000;
Kraemer et al. 2000), Mrk~3 (Collins et al. 2005, 2009), and NGC~1068
(Kraemer \& Crenshaw 2000a). These photoionisation models assume two or
three different components of ionised gas (at each distance from the nucleus),
photoionised directly by a power-law continuum and by an absorbed
continuum in the case of the low-ionisation component.
Similar models were used to reproduce high- and low-ionisation lines detected in the nuclei of
Mrk~573 (Kraemer et al. 2009) and NGC~1068 (Kraemer \& Crenshaw 2000b).

A selection of different models, including photoionisation models with a mix of matter- and ionisation-bounded clouds (Binette et al. 1996), standard power-law photoionisation models, and shock models (Dopita \& Sutherland 1996), are shown in Fig.~\ref{models}, together with the \NeV/[Ne\,{\sc iii}] and [O\,{\sc ii}]/\OIII\ line ratios measured for the galaxies in the sample. We have included the line ratios of NGC~4151, NGC~1068, and Mrk~3 for comparison purposes, but notice that these lines have already been modeleded by Nelson et al. (2000), Kraemer et al. (2000), Collins et al. (2005, 2009), and Kraemer \& Crenshaw (2000a). The models included in Fig.~\ref{models} were taken from fig.~4 of Tadhunter (2002). We also included a fiducial point from the dust-free models of Ferguson et al. (1997a) and those corresponding to the PB50 and BB20 models of Komossa \& Schulz (1997). These multi-component models are described in more detail below. Note that the photoionisation models shown in Fig.~\ref{models} do not aim at a detailed modelling of every single emission line, like \NeV, of every galaxy, but rather at  presenting global trends. 
From this diagnostic diagram we can see that the photoionisation models of Ferguson et al. (1997a) and Komossa \& Schulz (1997) reproduce the bulk of the line ratios fairly well. However, in some cases (e.g., Mrk~573 and NGC~1068) they underpredict the \NeV/[Ne\,{\sc iii}] ratio.
In a small parameter space, the shock plus precursor and photoionisation models overlap, and within a very narrow parameter space shock plus precursor models also match the data (Mrk~3, NGC~4151, and NGC~7682); while the bulk of these particular models are off-set from the data - they systematically overpredict [O\,{\sc ii}]/\OIII\ and underpredict [Ne\,{\sc iii}]/[Ne\,{\sc v}]. A similar result was obtained by Nelson et al. 2000 for NGC~4151.
Both, the pure shock models and the models including matter-bounded clouds overestimate the [O\,{\sc ii}]/\OIII\ line ratios, giving a poor description of the measured ratios in these galaxies. Although the line ratios presented here were not corrected for extinction, an extreme correction would be required for the [O\,{\sc ii}]/\OIII\ intrinsic ratio to lie in the regions predicted by these models.

It is interesting to note that the line ratios from these galaxies do not vary very much, neither from galaxy to galaxy nor within each object, occupying a small zone in the diagrams. This can be better seen in Fig.~\ref{U}, where we plot the [Ne\,{\sc iii}]/[Ne\,{\sc v}] and [O\,{\sc ii}]/\OIII\ ratios as a function of the distance to the nucleus of the galaxies. For most of the objects in our sample these ratios do not vary strongly with the distance and show very similar values. Since these ratios are usually interpreted as a measure of the ionisation parameter (e.g., Penston et al. 1990; Komossa \& Schulz 1997), their constancy would imply that the gas density in these galaxies decreases proportional to the square of the distance. 
Extra information concerning the extinction affecting the gas can be derived from these plots. As it was mentioned above, the line ratios were not corrected for extinction. However, the closeness in wavelength between the \NeV\ and [Ne\,{\sc iii}] lines makes its ratio almost independent of reddening. On the other hand, the [O\,{\sc ii}]/\OIII\ ratio is more affected by dust extinction. We can see in Fig.~\ref{U} that, in most cases, both ratios show a very similar behaviour with respect to the distance to the nucleus, suggesting that the emission-line gas is not strongly affected by dust extinction.

Next, we focus on just the CLs for which not all models shown in Fig.~\ref{models} are available. Given large uncertainties in the collision strengths of the iron
coronal lines (see, especially, the cautious comments by
Ferguson et al. 1997a), we limit any comparison between data
and models to order of magnitude estimates. 
Ferguson et al. (1997b) carried out photoionisation calculations to identify
the optimal conditions and locations in which the coronal lines form.
Assuming that radiation from the central AGN is the only excitation
mechanism, and using plane-parallel, constant density slabs of gas, they
determined the distances from the ionising source in which the coronal lines are
emitted as a function of density. They provide line equivalent widths in
the density-distance plane, indicating where the bulk of the emission occurs for
each line, and allowing rough comparisons between observed and
predicted emission-line flux ratios. The [Ne\,{\sc v}], [Fe\,{\sc vii}],
[Fe\,{\sc x}], [Fe\,{\sc xi}] and [Fe\,{\sc xiv}] lines observed here are
included in their calculations. They assumed an ionising EUV continuum similar to that of a typical Seyfert galaxy, with $L_{\rmn{ion}}=10^{43.5}$~erg~s$^{-1}$, where $L_{\rm{ion}}$ is the ionising luminosity of the central source.
These models predict that, under optimal conditions\footnote {log $U$(H) = $-1.25$, $-2$, $-0.5$, 0  and range of log $n$(H) = 2.0--7.0, 3.0--7.5, 3.0--8.25, 5.0--6.5 for \NeV, \FeVII, \FeX, and \FeXI, respectively.}, the radius of the region emitting the [Ne\,{\sc v}] and [Fe\,{\sc vii}] lines is between 0.4--130~pc and 0.6--100~pc,
respectively. For [Fe\,{\sc x}], they show that it is restricted to the
inner 20~pc while for \FeXI\ the emission region is not larger than one
parsec. 
These values agree with the size of the CLRs derived in this work in all objects but Mrk~3, NGC~1068, and Mrk~573, where moderate to large discrepancies are found. In the
former, both Fe$^{+6}$ and Ne$^{+4}$ extend to distances of about 200~pc.
For the latter two objects, the greatest discrepancy is for the lines of high IP, that is, [Fe\,{\sc x}] and \FeXI, which are detected to scales of $\sim 100$~pc and 60~pc, respectively. For NGC~1068, Pier et al. (1994) and Bland-Hawthorn et al. (1997) estimate ionising
luminosities 1--2 orders of magnitude higher than used by Ferguson et al. (1997b). If we take into account that the size of the emission region scales as $L_{\rm{ion}}^{1/2}$, adopting
$L_{\rm{ion}} = 10^{45}$~erg~s$^{-1}$ and leaving the other parameters of the model
constant, the sizes of the emission regions for the different lines are
increased by a factor of up to 4. The size of the region emitting \FeX\
is now in good agreement, but we now overestimate the size of the \FeVII\ region,
which should now be visible to scales of several hundred parsecs. Furthermore, the
region emitting \FeXI\ continues to be strongly underestimated. A similar
argument can be applied to Mrk~573, although in that source the region
containing \FeXI\ was not covered by STIS.

From a large grid of photoionisation models\footnote{With the same assumptions as in the Ferguson et al. (1997b) models, but the collision strengths of the iron transitions were set to 1 due to the large uncertainties in these quantities.},  
and assuming a ``locally optimally emitting clouds'' (LOC) scenario, Ferguson et al. (1997a) predicted integrated spectra for different distributions of cloud covering fraction and density. 
The integrated spectrum that best reproduces the [O\,{\sc ii}]/\OIII\ and \OIII/H$\beta$ ratios of a mean Seyfert~2 spectrum also predicts \FeX/\FeVII\ and \NeV/\FeVII\ ratios of 0.3 and 12.6, respectively. These values can be compared with the ones measured in the galaxies of the sample, although keeping in mind that Ferguson et al.'s  values are integrated over the whole NLR. The upper-left panel of Fig.~\ref{fe7fe10} shows the distribution of the \FeX/\FeVII\ line ratio found in NGC~1068, NGC~4151, and NGC~3227; the only galaxies where it could be measured. This ratio varies from about 0.1--1.2, with the largest values found at the nuclear positions. In the lower-left panel of Fig.~\ref{fe7fe10} we show the \NeV/\FeVII\ ratio, which varies from about 1--15. Additionally, we included in Fig.~\ref{fe7fe10} the ratios of the CLs over \OIII. In general, the line ratios tend to be higher at the nuclear positions, where the bulk of the CLs are emitted. 
We also added in Fig.~\ref{fe7fe10} the predicted ratios from the photoionisation models of Komossa \& Schulz (1997). Among their multi-component models, we representatively show two. These were computed for a mixture of gas densities and radii, varying the ionising EUV continuum: one continuum represented by a blackbody with $\rm{T}=200000$~K (BB20; for comparison purposes) and the other was described by a combination of a power law ($\alpha_{\rm{UV-X}}=-2$) and BB20, each contributing 50\% (PB50). 
It can be seen that the models of both, Ferguson et al. (1997a) and Komossa \& Schulz (1997),  do a relatively good job in reproducing the observed ratios, although they tend to underpredict the CLs measured at the most central regions of the galaxies. 
This is naturally expected, since the models included a mix over a range
of radii, while at the smallest nuclear distances the highest ionisation
lines are enhanced relative to the lower ionisation lines. Especially,
given their high critical densities,
coronal lines can be selectively boosted when adding a high-density
component (e.g., table~3 of Komossa \& Bade 1999).
Excluding the nuclear regions, models and observations typically agree within a factor of a few. 
Since CLs sensitively depend on the EUV-SED, employing higher-temperature EUV bump components will naturally increase the strengths of the CLs with respect to the most lower ionisation lines (Komossa \& Schulz 1997). Keeping further in mind the possibility of a contribution of matter-bounded clouds (which have a relatively smaller contribution from low-ionisation lines; e.g., Binette et al. 1997, Binette 1998) and the remaining uncertainties in the collisional strengths of the iron CLs (e.g., Ferguson et al. 1997b), we conclude that the consistency between photoionisation models and observations is gratifying.

In summary, the observational evidence presented in this section shows that
photoionisation by a central source can account remarkably well for most of the coronal and low-ionisation line ratios measured in the galaxies of the sample.

\subsection{Extinction: the case of NGC~1068}

Several lines of evidence suggest that strong extinction, at distances towards the SW of the nucleus of NGC~1068, is playing an important role in the shape of its NLR: optical and NIR emission lines and continuum emission are weaker towards the SW (e.g., Machetto et al. 1994; Kraemer \& Crenshaw 2000a; Thompson et al. 2001; Geballe et al. 2009); polarised light in the NIR towards the SW from the nucleus is not visible at optical wavelengths (Packham et al. 1997); and the silicate optical depth is strongly asymmetric, with a maximum at SSW position from the nucleus (Mason et al. 2006). 

The STIS observations analysed here also points to strong extinction affecting the emission at SW distances from the nucleus (see Section~\ref{ngc1068_results}). A close examination of Fig.~\ref{ngc1068_flux} shows that toward the SW most CLs are suppressed, \NeV\ being the only one that can be observed at some positions in that direction. Only \OIII, the brightest of all optical lines, is visible at all apertures but its strength, compared to that of the NE side, is strongly suppressed, by up to a factor of 10. This suggests that indeed, strong extinction is present to the SW. The nucleus itself seems to be strongly attenuated too, as can be observed from the distribution of the ratio \NeV/\FeVII\ along the spatial direction measured in NGC~1068 shown in Fig.~\ref{fe7fe10}.  This ratio is strongly sensitive to reddening because of the large difference in wavelength of the lines involved. It is independent of the form of the ionising continuum and likely of chemical abundances (except if metallicity shows a strong gradient along the slit).
Fig.~\ref{fe7fe10} shows that from about 20~pc NE of the nucleus and outwards, the ratio \NeV/\FeVII\ is relatively constant, with an average value of $7\pm2$. In contrast, from 20~pc NE to the SW, it drops sharply, reaching 0.5 at only 25~pc to the SW. Moreover, Fig.~\ref{ngc1068_flux} tells us that \NeV\ follows exactly the same spatial distribution as \OIII. No \NeV\ is detected at 50~pc SW, where a minimum in the \OIII\ distribution is observed. A few positions outwards, where the \OIII\ emission increases again, \NeV\ also does, becoming visible up to $\sim$120~pc, where it is no longer detected. Note that to the NE, the maximum distance where we detect \NeV\ is 140~pc. We therefore confirm that the size of the \NeV\ emission region measured to the NE is very similar to that towards the SW. We conclude, therefore, that it is the strong extinction to the SW that blocks the detection of CLs in this direction. Note that the dust blocking the coronal lines is not spatially uniform. Rodr\'{\i}guez-Ardila et al. (2006) report the detection of \FeVII\ up to 120~pc south of the nucleus. In their observations, these authors employed a 1$\arcsec$ slit aligned in the North-South direction.

It calls the attention that the coronal lines that seem to be strongly extinguished along the STIS slit in NGC~1068 are those coming from ions that are usually locked up into grains: iron and silicon. Therefore, in summary, dust extinction, and the gas phase depletion of refractory elementens, very plausibly explain the observations of NGC~1068.

\section{Summary and conclusions}

We have presented a study of the CLR of a sample of ten Seyfert galaxies based on optical STIS/HST spectra. These spectra allowed us to resolve the regions emitting CLs, and to analyse the properties of the CLs as a function of the distance to the nucleus. Our main results can be summarised as follows:

\begin{enumerate}

\item CLs display very similar flux distributions than low-ionisation lines, showing a clumpy morphology. The CLRs extent from less than 10~pc up to 230~pc in radius; the most
compact one is observed in NGC~3227, the most extended one in Mrk~3. We confirm, that high-ionisation CLs arise from more compact regions than the lower ionisation CLs.
The compactness of the CLR is consistent with an origin of the bulk of the CLs between BLR and NLR, and extending well into the NLR (up to 100s of pc) in \FeVII\ and \NeV.
Moreover, the distribution of the ratios between CLs and \OIII\ shows its maximum at the nucleus.

\item The two (ionisation-parameter sensitive) line ratios, [O\,{\sc ii}]/\OIII\ and [Ne\,{\sc iii}/\NeV, generally show very similar radial dependencies.

\item The highest ionisation CLRs appear to be the ones in NGC~4151 and NGC~3227, in terms of the \FeX/\FeVII\ line ratio. However, only in NGC~1068 we observed [Si\,{\sc xii}], the highest ionisation line observed in the galaxies of the sample.  

\item In general, the CL profiles show strong asymmetries, that vary with the distance to the nucleus. In some cases we observe line splitting in the core, sometimes locally complex with
several components contributing. In particular, a remarkable double-peak structure was observed in the \FeX\ lines of Mrk~573 and NGC~4507. In the case of
Mrk~573, the relative separation between the two peaks increases when going
from NW to SE. This signals the presence of a highly energetic outflow,
not detected in lower ionisation lines.

\item Patterns of rotation curves in \OIII\ are generally followed by CLs. Local
deviations occur, occasionally. Maximal $\Delta$V of CLs reach up to
typically 500\kms\ (1000\kms\ in case of NGC~1068) with both,
red- and blueshifts occurring. 
Variation in width and peak position of the lines occurs from point to
point without any particular universal trend with respect to the IP of the lines.

\item The presence or absence of CLs, their strengths, and kinematics, does not
scale in any obvious and strong way with the radio properties (position of radio knots in jets).

\item Several lines of evidence point towards photoionisation as major ionisation mechanism: the generally close spatial dependencies between CLs and low-ionisation lines,
the low inferred gas temperatures of order 10000--20000~K
typical of photoionised gas, and the fact that available photoionisation
models are generally successful in matching observed line ratios,
within a factor of a few.  

\end{enumerate}

Our work demonstrated the power of spatially resolved spectroscopy to provide information of the inner parts of AGNs, where the bulk of CLs arise. A full comprehension of these lines is very important in order to improve our knowledge about energetic processes, such as outflows and shocks, occurring in the very galaxy cores.

\section*{Acknowledgments}

XM acknowledges scholarships from the Deutscher Akademischer Austausch Dienst (DAAD) and from the Consejo Nacional de Investigaciones Cient\'ificas y T\'ecnicas (CONICET). A.R.A acknowledges support of the Brazilian Funding Agency CNPq under grant
311476/2006-6.

\end{document}